\documentclass[a4paper,12pt]{article}
\usepackage[utf8]{inputenc}
\usepackage[english]{babel}
\usepackage{amsfonts}
\usepackage{latexsym,amsmath,amssymb,amsthm}
\usepackage[active]{srcltx}
\usepackage{hyperref}
\usepackage[all]{xy}
\input diagxy
\usepackage{tikz-cd}
\font\fr=eufm10 scaled \magstep 1 
\newtheorem{teor}{Theorem}

\newtheorem{definition}{Definition}
\theoremstyle{definition}

\theoremstyle{remark}

\newtheorem{remark}{Remark}
\def\beq{\begin{equation}}
\def\eeq{\end{equation}}
\def\bea{\begin{eqnarray}}
\def\eea{\end{eqnarray}}
\def\beann{\begin{eqnarray*}}
\def\eeann{\end{eqnarray*}}
\def\beasn{\begin{sneqnarray}}
\def\eeasn{\end{sneqnarray}}
\def\ben{\begin{enumerate}}
\def\een{\end{enumerate}}
\def\bit{\begin{itemize}}
\def\eit{\end{itemize}}
\def\derpar#1#2{\frac{\partial{#1}}{\partial{#2}}}

\def\qed{\ifvmode\removelastskip\fi
{\unskip\nobreak\hfil\penalty50\hbox{}\nobreak\hfil
\hbox{\vrule height1.2ex width1.2ex}\parfillskip=0pt
\finalhyphendemerits=0 \par\smallskip}}
\def\buit{\hbox{\rm\O}}
\def\vf{\mbox{\fr X}}
\def\df{{\mit\Omega}}
\def\Lag{{\cal L}}

\def\d{{\rm d}}

\def\Real{\mathbb{R}}

\def\Tan{{\rm T}}
\def\Lie{\mathop{\rm L}\nolimits}
\def\inn{\mathop{i}\nolimits}
\def\Cinfty{{\rm C}^\infty}

\renewcommand{\neq}{=\hspace{-3.5mm}/\hspace{2mm}}

\newcommand*{\dd}{\mathrm{d}}

\title{\sc Constrained Lagrangian dissipative contact dynamics}%Herglotz variational problem with constraints}
\author{\sffamily 
\sc $^a$Manuel de Le\'on\thanks{mdeleon@icmat.es\,({\it ORCID}:\,0000-0002-8028-2348).}\ ,
$^b$Manuel La\'inz\thanks{manuel.lainz@icmat.es\,({\it ORCID}:\,0000-0002-2368-5853).}\ , \\
\sc $^c$Miguel C. Mu\~noz-Lecanda\thanks{miguel.carlos.munoz@upc.edu\,({\it ORCID}:\,0000-0002-7037-0248).}\ ,
$^c$Narciso Rom\'an-Roy\thanks{narciso.roman@upc.edu\,({\it ORCID}:\,0000-0003-3663-9861).}\ .
\\[1ex]
\normalsize\itshape\sffamily 
$^a$Instituto de Ciencias Matem\'aticas,
Consejo Superior de Investigaciones Cient\'ificas\\
\normalsize\itshape\sffamily 
and Real Academia de Ciencias, Madrid, Spain.
\\[1ex]
\normalsize\itshape\sffamily 
$^b$Instituto de Ciencias Matem\'aticas,
Consejo Superior de Investigaciones Cient\'ificas, Madrid, Spain.
\\[1ex]
\normalsize\itshape\sffamily 
$^c$Department of Mathematics,
Universitat Polit\`ecnica de Catalunya,
Barcelona, Spain.
}

\date{\today} 
\begin{document}
\maketitle
\begin{abstract}
We show that the contact dynamics obtained from the Herglotz variational principle can be described as a constrained nonholonomic or vakonomic ordinary Lagrangian system depending on a dissipative variable with an adequate choice of one constraint. As a consequence we obtain the dynamics of contact nonholonomic and vakonomic systems as ordinary variational calculus with constraints on a Lagrangian with a dissipative variable. The variation of  the energy and the other dissipative quantities are also obtained giving the usual results.
\end{abstract}


\medskip
\vbox{\raggedleft AMS s.\,c.\,(2020): 
{\it Primary\/}: 37J55, 53D10, 70G75.
{\it Secondary\/}: 37J06, 70G45, 70H03.}\null

\markright{{\rm M. de Le\'on} {\it et al\/},
    {\sl Constrained Lagrangian dissipative contact dynamics.}}

\newpage
%\medskip
\setcounter{tocdepth}{2}
{
\def\baselinestretch{0.97}
\small
% hack per a eliminar l'espai vertical 1em
\def\addvspace#1{\vskip 1pt}
\parskip 0pt plus 0.1mm
\tableofcontents
}

%%%%%%%%%%%%%%%%%%%%%%%%%%%%%%%%%%%%%%%%%
%%%%%%%%%%%%%%%%%%%%%%%%%%%%%%%%%%%%%%%%%%%
%%%%%%%%%%%%%%%%%%%%%%%%%%%%%%%%%%%%%%%%%%%

\section{Introduction}

The symplectic formulation of Hamiltonian mechanics is well known, and has allowed us important developments since the 1960s
\cite{AM-78,Ar-89,LM-sgam}. 
The Lagrangian formalism accompanied the Hamiltonian one during this time. 
More recently, the study of the geometrization of dissipative systems \cite{CG-2019,Galley-2013,MPR-2018},
in its natural framework of contact geometry \cite{Banyaga2016,Geiges2008,KA-2013}, 
 has gained momentum, although the Lagrangian version was available since the 1930s thanks to the developments of Gustav Herglotz
 \cite{He-1930,Her-1985}. 
Contact mechanics 
\cite{Bravetti2017,BCT-2017,BLMP-2020,CIAGLIA2018,LGLMR-2021,LGMMR-2020,LGMMR-2020,DeLeon2019,GGMRR-2019b,LL-2018}
is also the natural framework for studying thermodynamics as early as Constantin Carath\'eodory 
(the first quarter of the past century), although its applications today cover many other fields
\cite{Bravetti2018,CIAGLIA2018,GGMRR-2019,GGMRR-2020,Goto-2016,KA-2013,RMS-2017,Ra-2006}.

In this paper we show how these versions (including also the cases of the noholonomic and vakonomic mechanics) 
can be considered under a common scenario, making a proper use of the so-called presymplectic geometry and the Euler-Lagrange operator. 
One of the most interesting contributions of this paper is precisely the simplicity of this treatment, which allows us also to compare both dynamics in a very clear way.
After recalling the well-known descriptions of Lagrangian and Hamiltonian mechanics in the usual case (based on the symplectic formalism and Hamilton's principle), 
we proceed to consider the case that constitutes the essential part of this paper.

We start with a ``dissipative'' regular Lagrangian $\mathfrak{L}\colon\Tan Q\times\Real\longrightarrow \mathbb{R}$;
that is, $\mathfrak{L}=\mathfrak{L}(q,v,s)$, and extend it to a singular Lagrangian
$L\colon\Tan(Q\times\mathbb{R})\longrightarrow\mathbb{R}$. 
Then, we consider the problem of finding the dynamics for a presymplectic system determined on $\Tan(Q\times\mathbb{R})$ and subject to the constraint $v_s=L$, where $v_s=\dot{s}$ as geometric coordinate in $\Tan (Q\times\Real)$. 
Whether we consider the nonholonomic or the vakonomic formalisms, 
if we develop the presymplectic algorithm, we obtain the same dynamics, which is none other than the Herglotz dynamics.

The situation is quite different in the case when, in addition to the constraint $v_s=L$, 
we consider the existence of additional constraints in $\Tan Q$; 
that is, provided by a submanifold $N$ of $\Tan Q$. 
In this case, we must take into account all the constraints
(defining a submanifold $C$ of $N$, where $N$ is considered now
as a submanifold in $\Tan(Q\times \mathbb{R})$ in the obvious way)
and we obtain different equations which we can call the dissipative 
{\sl nonholonomic and vakonomic equations}, respectively. 
We also note that, if the constraints defining $N$ 
do so from a submanifold of $\Tan Q\times\mathbb{R}$, 
in the nonholonomic case we obtain the same equations as when there is no dependence on $s$, 
while in the vakonomic case
we recover the equations obtained in \cite{LLM-2021}.

The paper is structured as follows. In section 2, we recall the main geometric and dynamical elements of Lagrangian and Hamiltonian mechanics, including the cases of noholonomic and vakonomic mechanics, and making use of the Euler-Lagrange operator, which will play a crucial role in the rest of the paper. 
At the end of the section, we include the Herglotz variational principle, and its relation to the description of the mechanics in terms of the contact geometry. 
Section 3 is devoted to the study of the nonholonomic and vakonomic equations and how they can be obtained by using the presymplectic formalism and the constraint $\dot{s}=L$, 
where $s$ is the standard coordinate (the dissipation) and $L$ is the Lagrangian. 
In both cases, the result obtained is the same: the Herglotz equations. 
The situation changes when, in addition to this dynamical constraint, we have a submanifold of constraints. 
These cases are analyzed in sections 4 and 5 and we find that the results differ: there are two dynamics which in general do not coincide, the nonholonomic and the vakonomic dynamics.

All the manifolds are real, second countable and $\Cinfty$.
Manifolds and mappings are assumed to be smooth.
All the differential forms are supposed to have constant rank.
Sum over crossed repeated indices is understood.

%%%%%%%%%%%%%%%%%%%%%%%%%%%%%%%%%%%%%%%%%%%%%%%%%%%%%%%%%%%%%%%%%%

\section{Preliminaries}

\subsection{Geometric Lagrangian mechanics}

First we present a brief review about the 
{\sl Klein} or {\sl Cartan formulation} of the Lagrangian formalism 
\cite{Cr-83,CP-adg,dLe89,klein,SCC-84}.

Let $Q$ be an $n$-dimensional manifold representing the
{\sl configuration space} of a dynamical system.
In the Lagrangian formalism, the {\sl velocity phase space}
of the system is represented by the tangent bundle 
$\pi_Q\colon\Tan Q\to Q$.
Then, given a Lagrangian function $\Lag\in\Cinfty(\Tan Q)$,
we can use the canonical structures of $\Tan Q$
(the vertical endomorphism $J \colon\vf(\Tan Q)\to\vf(\Tan Q)$
and the Liouville vector field $\Delta\in\vf(\Tan Q)$)
to construct the \textsl{Lagrangian energy} associated with $\Lag$
$$
E_\Lag := \Delta (\Lag ) - \Lag \in \Cinfty (\Tan Q) \ ,
$$
and the \textsl{Lagrangian forms} associated with $\Lag$
$$
\theta_\Lag := \inn_J\d\Lag=\d \Lag \circ J \in \df^1(\Tan Q) 
\quad ; \quad
\omega_\Lag := -\d \theta_\Lag \in \df^2(\Tan Q) \ ,
$$
where $\df^k(\Tan Q)$ denotes the set of differentiable $k$-forms in $(\Tan Q)$.
In a a natural chart of natural coordinates $(q^i,v^i)$ in $\Tan Q$,
the local expression of the vertical endomorphism and 
the Liouville vector field are
$\displaystyle J= \d q^i\otimes \derpar{}{v^i}$ and
$\displaystyle\Delta= v^i\derpar{}{v^i}$; then,
given a Lagrangian function $\Lag=\Lag (q^i,v^i)$,
the local expressions of the action and the Lagrangian energy associated with the Lagrangian $\Lag$ are 
$$
E_\Lag = v^i\derpar{\Lag}{v^i} - \Lag \quad , \quad
\theta_\Lag= \derpar{\Lag}{v^i}\,\d q^i \quad , \quad
\omega_\Lag=
\frac{\partial^2\Lag}{\partial q^j \partial v^i}\, \d q^i \wedge \d q^j +
\frac{\partial^2\Lag}{\partial v^j \partial v^i}\, \d q^i \wedge \d v^j \ .
$$
We say that
$\Lag \in \Cinfty (\Tan Q)$ is a  \textsl{regular Lagrangian function}
(and that $(\Tan Q,\Lag )$ is a \textsl{regular Lagrangian system})
if $\omega_\Lag$ is nondegenerated;
otherwise, $\Lag$ is a  \textsl{singular Lagrangian function}
(and $(\Tan Q,\Lag )$ is a {\sl singular Lagrangian system}).
Locally, $\Lag$ is a regular Lagrangian if, and only if,
in a natural coordinate chart of $\Tan Q$, the partial Hessian matrix
\(\displaystyle {\cal H}_\Lag:=
(\frac{\partial^2\Lag}{\partial v^j \partial v^i})\)
is regular at every point
and then $\omega_\Lag$ is a symplectic form.

The dynamical trajectories of a Lagrangian system
$(\Tan Q,\Lag)$ are the integral curves
of a vector field $X_\Lag\in\vf(\Tan Q)$ such that
it is a solution to the {\sl Lagrangian equation}
\beq
\inn_{X_\Lag}\omega_\Lag = \d E_\Lag \ ,
\label{elm}
\eeq
and it is a {\sl second-order differential equation} {\sc (sode)}; 
that is, 
\beq
J (X_\Lag ) = \Delta \ .
\label{edso}
\eeq
A vector field $X_\Lag$ solution to \eqref{elm} (if it exists) is 
said to be a \textsl{Lagrangian vector field} and,
if the condition (\ref{edso}) holds, then
it is called an \textsl{Euler-Lagrange vector field}. 
If $X_\Lag$ is a {\sc sode}, then its integral curves are {\sl holonomic}
(i.e.; canonical liftings to $\Tan Q$ of curves in $Q$).

In natural coordinates of $\Tan Q$,
if
\(\displaystyle X_\Lag= A^i\derpar{}{q^i}+B^i\derpar{}{v^i}\),
as
$$
\d E_{\cal L}= v^h\frac{\partial^2 L}{\partial q^i\partial v^h} \d q^i+
v^h\frac{\partial^2{\cal L}}{\partial v^i\partial v^h}\d v^i-
\frac{\partial{\cal L}}{\partial q^i}\d q^i\ ,
$$
equation (\ref{elm}) leads to
\bea
\frac{\partial^2\Lag}{\partial v^j \partial v^i}B^i -
(\frac{\partial^2\Lag}{\partial q^j \partial v^i} +
\frac{\partial^2\Lag}{\partial v^j \partial q^i})A^i +
\frac{\partial^2\Lag}{\partial v^i \partial q^j}v^i -
\derpar{\Lag}{q^j} &=& 0 \ ,
\label{eq11a}
\\
\frac{\partial^2\Lag}{\partial v^j \partial v^i}(A^i-v^i) &=& 0 \ .
\label{eq11}
\eea
As condition (\ref{edso}) is locally equivalent to
$A^i=v^i$ and the integral curves $\sigma\colon\Real\to\Tan Q$ 
of a {\sc sode} $X_\Lag$ are canonical liftings of curves
$\gamma\colon\Real\to Q$; then, if $\gamma(t)=(q^i(t))$ and 
$\sigma(t)=(q^i(t),\dot q^i(t))$, we have that
$$
A^i = v^i = \frac{d q^i}{d t}
\quad , \quad
B^i = \frac{d^2q^i}{d t^2} \ ,
$$
and these expressions together with (\ref{eq11a}) and (\ref{eq11}) lead to 
\beq
(\frac{\partial^2\Lag}{\partial v^j \partial v^i}\circ\sigma)
\frac{d^2q^i}{d t^2}-
(\frac{\partial^2\Lag}{\partial v^j \partial q^i}\circ\sigma)
\frac{d q^i}{d t} +
\derpar{\Lag}{q^j}\circ\sigma=
\frac{d}{d t}(\derpar{\Lag}{v^j}\circ\sigma)-
\derpar{\Lag}{q^j}\circ\sigma=0 \ ;
\label{ELequats}
\eeq
which are the {\sl Euler-Lagrange equations} for the dynamical trajectories.

Therefore, if $(\Tan Q,\Lag)$ is a regular Lagrangian system,
as a consequence of the regularity of the Hessian matrix ${\cal H}_\Lag$,
there exists a unique vector field $X_\Lag\in\vf(\Tan Q)$
which is the solution to the Lagrangian equation
\eqref{elm} and it is a {\sc sode}.

\begin{remark}
If the Lagrangian system is singular, then equation \eqref{elm}
is not necessarily compatible everywhere on $\Tan Q$  and, 
when it has solution,
it is not unique and it is not necessarily a {\sc sode};
so the condition \eqref{edso} must be added to the Lagrangian equations
in order to obtain the Euler--Lagrange equations \eqref{ELequats}.
In general, solutions $X_\Lag$ exist only in some submanifold $S_f\hookrightarrow\Tan Q$,
and a {\sl constraint algorithm} must be implemented in order to find 
this submanifold $S_f$ (if it exists) where there are 
{\sc sode} vector fields $X_{\cal L}\in\vf(\Tan Q)$,
tangent to $S_f$, which are solutions to the Lagrangian equations on $S_f$.
The guidelines of this algorithm are the following:
\begin{itemize}
\item
First we find the {\sl compatibility conditions}
which define the submanifold of $\Tan Q$ where the
Lagrangian equation \eqref{elm} is compatible.
This submanifold is defined by
$$
P_1=\{ {\rm p}\in\Tan Q\, \vert\, (\inn_Z\d E_{\cal L})(p)=0\, ,
\ \mbox{\rm for every $Z\in\ker\,\omega_{\cal L}$} \}\ .
$$
\item
Then we apply the {\sl tangency conditions}:
From a local basis of vector fields $Z\in\ker\,\omega_{\cal L}$
we obtain a basis of independent constraint functions
$\zeta^Z=\inn_Z\d E_{\cal L}$ locally defining $P_1$.
Then, if $X_{\cal L}\in\vf(\Tan Q)$ is the general solution to  \eqref{elm} on $P_1$, these tangency conditions are
$$
(\Lie_{X_{\cal L}}\zeta^Z)\vert_{P_1}=0 \ .
$$
These conditions can originate new constraints defining a submanifold
$P_2\hookrightarrow P_1$ or not. In the first case,
this step is iterated until no new constraints appear;
in which case we obtain a final submanifold $P_f\hookrightarrow\ldots\hookrightarrow P_1\hookrightarrow\Tan Q$;
or until $P_f=\buit$, or $P_f$ is a set of isolated points.
\item
{\sl Second-order condition}:
As $X$ could not be a {\sc sode}, if we impose the condition $J(X)=\Delta$ 
to the general solution $X\in\vf(\Tan Q)$,
new constraints can appear defining
a submanifold $S_1\hookrightarrow P_1$.
Therefore, the tangency conditions on these new constraints 
can originate new constraints which,
at the end of the algorithmic procedure
(in the best of cases) lead to obtain the final submanifold $S_f\hookrightarrow P_f$.
\end{itemize}
(See \cite{BGPR-86,CLR-88,GN-79,GN-80,MR-92} for details
on these topics).
\end{remark}

\subsection{The Euler--Lagrange operator}

In the variational formulation of classical mechanics, the dynamical equations are obtained from the Hamilton Principle: Given a Lagrangian function ${\cal L}\in\Cinfty(\Tan Q)$ we try to find the curves $\gamma:I\subset\Real\to Q$ being critical points of the functional
$$
\gamma\mapsto \int_I {\cal L}\circ\dot\gamma \,,
$$
and we obtain the Euler--Lagrange equations.

From a more intrinsic point of view we can introduce the Euler--Lagrange operator which is a very relevant tool in geometric Lagrangian mechanics
(see \cite{Tu-74,Tu-76a,Tu-76b} for the definition and applications). We give here a brief approach in order to be used freely in the sequel.

For a Lagrangian ${\cal L}\in\Cinfty(\Tan Q)$ the Euler--Lagrange operator is defined as follows:
Let $\d^V{\cal L}\colon\Tan Q\to\Tan^*Q$ be the fibre derivative of ${\cal L}$
(whose local expression is 
$\displaystyle\d^V{\cal L}(q^i,v^i)=\Big(q^i,\derpar{L}{v^i}\Big)$); 
denote by ${\mathbf{s}}\colon\Tan(\Tan Q)\to\Tan(\Tan Q)$ the {\sl canonical involution}
(which is an isomorphism whose local expression is 
$\mathbf{s}(q,v;u,a)=(q,u;v,a)$, for $(q,v;u,a)\in\Tan(\Tan Q)$); and, finally,
denote by $\displaystyle D\equiv\frac{d}{dt}$
the {\sl total time-derivative operator}; then:

\begin{definition}
\label{EL} 
The \textbf{Euler-Lagrange operator} associated
with the Lagrangian ${\cal L}$ is the map 
${\mathcal{E}}_{\cal L}\colon \Tan^2Q\to\Tan^*Q$ defined by 
$$
\langle\mathcal{E}_{\cal L}\circ\ddot{\gamma},\mathbf{w}\rangle
:=\langle\d {\cal L}\circ\dot{\gamma},\mathbf{s}\circ\mathbf{\dot w}\rangle-
\mathrm{D}\langle\d^\mathrm{v}{\cal L}\circ\gamma,\mathbf{w}\rangle \ ;
$$
for every path $\gamma\colon I\subset\Real\to Q$ and
every vector field $\bf w$ along~$\gamma$
(i.e., $\mathbf{w}\colon I\subset\Real\to\Tan Q$
such that $\pi_Q\circ\mathbf{w}=\gamma$),
which are representatives of points $v\in\Tan_q Q$
and $A=(q,u,a)\in\Tan^2Q$.
\end{definition}

Thus, the Euler-Lagrange operator is a one-form along the projection
$\Tan^2Q\to Q$, and it is also an affine bundle map along
the projection  $\Tan Q\to Q$. 
In this way, $\langle\mathcal{E}_{\cal L}(A),v\rangle$ is well defined by the above expression and, in local natural coordinates, we have that
\beq
\label{ELop}
\mathcal{E}_{\cal L}=(\frac{\partial{\cal L}}{\partial q^i}-\frac{\d}{\d t}(\frac{\partial{\cal L}}{\partial v^i}))\d q^i \ .
\eeq
Observe also that, if $f\colon\Real\to\Real$, then
\begin{equation*}
\mathcal{E}_{f{\cal L}}= f\mathcal{E}_{\cal L}-\frac{\d f}{\d t}\;\d^\mathrm{v}{\cal L} \ .
\end{equation*}

An important result concerning this operator is that
%\begin{prop}
a {\sc sode} $X\in\vf(\Tan Q)$ is solution to the equation 
$\inn_{X}\omega_{\cal L}-\d E_{\cal L}=0$ if, and only if, 
its integral curves $\gamma$ are solution to the equation 
$\mathcal{E}_{\cal L}\circ\ddot{\gamma}=0$.
%\end{prop} \proof
In fact,  in local natural coordinates we have that
$\displaystyle \omega_{\cal L}=\d q^i\wedge\d(\frac{\partial{\cal L}}{\partial v^i})$;
and, if
$\displaystyle 
X=v^j\frac{\partial }{\partial q^j}+B^j\frac{\partial }{\partial v^j}$,
then
\begin{eqnarray*}
\inn_{X}\omega_{\cal L}&=&v^i\d(\frac{\partial{\cal L}}{\partial v^i})
-\Lie_X(\frac{\partial{\cal L}}{\partial v^i})\d q^i\\[2mm]
&=&
v^h\frac{\partial^2{\cal L}}{\partial q^i\partial v^h} \d q^i+
v^h\frac{\partial^2{\cal L}}{\partial v^i\partial v^h}\d v^i
-v^j\frac{\partial^2{\cal L}}{\partial q^j\partial v^i}\d q^i-B^j\frac{\partial^2{\cal L}}{\partial v^j\partial v^i}\d q^i\ ,
\end{eqnarray*}
hence:
\begin{eqnarray}\label{dynvf=E-Lop}
\inn_{X}\omega_{\cal L}-\d E_{\cal L}&=&\frac{\partial{\cal L}}{\partial q^i}\d q^i
-v^j\frac{\partial^2{\cal L}}{\partial q^j\partial v^i}\d q^i-B^j\frac{\partial^2{\cal L}}{\partial v^j\partial v^i}\d q^i\nonumber\\
&=&
(\frac{\partial{\cal L}}{\partial q^i}
-v^j\frac{\partial^2{\cal L}}{\partial q^j\partial v^i}-B^j\frac{\partial^2{\cal L}}{\partial v^j\partial v^i})\d q^i=
(\frac{\partial{\cal L}}{\partial q^i}-\frac{\d}{\d t}(\frac{\partial{\cal L}}{\partial v^i}))\d q^i
\, ,
\end{eqnarray}
and taking into account that $B^j=\dot{v}^j$, this leads to the same expression than \eqref{ELop}.
%\qed

Hence we conclude that for a curve $\gamma$, to satisfy Euler-Lagrange equations for the Lagrangian ${\cal L}$ is equivalent to say that $\mathcal{E}_{\cal L}\circ\ddot\gamma=0$. 

As a final remark, consider the case of a product manifold
$M=Q_1\times Q_2$, hence $	\Tan M=\Tan(Q_1\times Q_2)=\Tan Q_1\times_{Q_1\times Q_2} \Tan Q_2$.
For a function $F\in\Cinfty(M)$, the product $M=Q_1\times Q_2$ provides a decomposition of
the Euler--Lagrange operator
$$
\mathcal{E}_F=\mathcal{E}_F^{Q_1}+\mathcal{E}_F^{Q_2}\,.
$$
In local coordinates $(q_1^i,q_2^j)$ of $Q_1\times Q_2$, the expression of this decomposition is
$$
\mathcal{E}_F=\mathcal{E}_F^{Q_1}+\mathcal{E}_F^{Q_2}=
(\frac{\partial F}{\partial q_1^i}-\frac{\d}{\d t}(\frac{\partial F}{\partial v_1^i}))\d q_1^i
+(\frac{\partial F}{\partial q_2^j}-\frac{\d}{\d t}(\frac{\partial F}{\partial v_2^j}))\d q_2^j\, ,
$$
and, for the operator $\d^\mathrm{v}$, we have
$$
\d^\mathrm{v}F=\d_1^\mathrm{v}F+\d_2^\mathrm{v}F=\frac{\partial F}{\partial v_1^i}\d q_1^i+\frac{\partial F}{\partial v_2^j}\d q_2^j\ .
$$
In the particular case that  $M=Q\times\Real$, with local coordinates $q^i,s)$, we have
\begin{equation}\label{E-L-product}
\mathcal{E}_F=\mathcal{E}_F^{Q}+\mathcal{E}_F^{\Real}=
(\frac{\partial F}{\partial q^i}-\frac{\d}{\d t}(\frac{\partial F}{\partial v^i}))\d q^i
+(\frac{\partial F}{\partial s}-\frac{\d}{\d t}(\frac{\partial F}{\partial \dot{s}}))\d s\,,
\end{equation}
and for the operator $\d^\mathrm{v}$,
\begin{equation}\label{d^v-product}
\d^\mathrm{v}F=\d_Q^\mathrm{v}F+\d_\Real^\mathrm{v}F=\frac{\partial F}{\partial v^i}\d q^i+\frac{\partial F}{\partial \dot{s}}\d s\ .
\end{equation}

\subsection{Nonholonomic Lagrangian mechanics}

{\sl Nonholonomic mechanics} describes dynamical systems
subjected to constraints depending on positions and velocities
\cite{Ar-89,CLMM-2002,LM-96,Ga,Lewis2020,Ve}.

The geometric description consists in taking a submanifold 
$C\hookrightarrow\Tan Q$ of codimension~$m<n=\dim Q$ and such that $\tau_Q(C)=Q$. 
Locally $C$ is defined by the vanishing of a set of constraints 
$\phi^i\in\Cinfty(\Tan Q)$ ($i=1,\dots ,m$)
such that $\d\phi^i$ are linearly independent at each point of $C$.
The projection $C \longrightarrow Q$ is assumed to be a submersion, 
and this is equivalent to take the constraints $\phi^i$ 
such that their fibre derivatives $\d^V\phi^i$ 
are linearly independent at every point of $C$. 
In local coordinates this means that the matrix 
$\displaystyle\Big(\derpar{\phi^i}{v^k}\Big)$ has maximal rank; that
is, the constraints restrict the velocities, not the positions.

An {\sl admissible path} is a curve 
$\gamma\colon I\subset\Real\to Q$ such that
$\dot\gamma$ takes its values in $C$.
Then, given a Lagrangian function $\Lag\in\Cinfty(\Tan Q)$, the triple 
$(\Tan Q,\Lag,C)$ is said to be a {\sl nonholonomic Lagrangian system}.
The {\sl nonholonomic variational problem} for this system consists in looking
for the admissible paths $\gamma$ which are solution to the equation
\beq
\mathcal{E}_\Lag\circ\ddot\gamma =
\lambda_i\,\d^V\phi^i \circ \dot\gamma \ ,
\label{nhELo}
\eeq
for some functions $\lambda_i\in\Cinfty(\Tan Q)$.
This variational problem can be stated in an equivalent way
looking for vector fields $X_\Lag\in\vf(\Tan Q)$
satisfying the following conditions:

(i) $X_\Lag$ is a {\sc sode}.

(ii) $X_\Lag$ is tangent to $C$.

(iii) $X_\Lag$ is solution to the equation:
\begin{equation}
    \inn_{X_\Lag}\omega_\Lag-\d E_\Lag=\lambda_i\d^\mathrm{v}\phi^i\, .
    \label{holec}
\end{equation}
If $X_\Lag$ is a {\sc sode} solution to this last  equation,
then its integral curves $\gamma$ are admisible paths;
that is, they are the solutions to \eqref{nhELo} which,
in natural coordinates, read
$$
\frac{d}{d t}(\derpar{\Lag}{v^j}\circ\dot\gamma)-
\derpar{\Lag}{q^j}\circ\dot\gamma=\lambda^i\derpar{\phi^i}{v^j}\circ\dot\gamma \ .
$$ 
These are the {\sl Euler--Lagrange equations of nonholonomic mechanics}.

One can easily verify that, if the partial Hessian matrix of $\Lag$ with respect to the velocities is positive or negative
definite, then there exists a unique solution to equation \eqref{holec}
(see \cite{LM-96} and \cite{LJL-2021} for the presymplectic and the contact cases, respectively).

%%%%%%%%%%%%%%%%%%%%%%%%%%%%%%%%%%%%%%%%%

\subsection{Vakonomic Lagrangian mechanics}

{\sl Vakonomic mechanics} is the treatment of constrained systems
which consists in modifying the variational principle
considering only the variations allowed by the constraints
(see, for instance, \cite{Ar-89,Bli,CLMM-2002,LMM,Els,MCL-2000} for details).

Now, the {\sl vakonomic variational problem} for this system consists in looking
for the admissible paths $\gamma$ which are solution to the equation
\beq
\mathcal{E}_{\Lag+\mu_i\phi^i}\circ\ddot\gamma=0 \ .
\label{vELo}
\eeq
for some functions $\mu_i\colon I\subset\Real\to\Real$.
As in the above case, this variational problem can be stated in an equivalent way
looking for vector fields $Y_\Lag\in\vf(\Tan Q)$
satisfying the following conditions:

(i) $Y_\Lag$ is a {\sc sode}.

(ii) $Y_\Lag$ is tangent to $C$.

(iii) $Y_\Lag$ is solution to the equation:
$$
\inn_{Y_\Lag}\omega_\Lag-\d E_\Lag=\mathcal{E}_{\mu_i\psi^i}=\dot{\mu}_i\d^\mathrm{v}\psi^i+\mu_i\mathcal{E}_{\psi^i}\ .
$$
If $Y_\Lag$ is a {\sc sode} solution to this last equation,
then its integral curves $\gamma$ are admisible paths;
that is, they are the solutions to \eqref{vELo} which,
in natural coordinates, read
$$
\frac{d}{d t}(\derpar{\Lag}{v^j}\circ\dot\gamma)-
\derpar{\Lag}{q^j}\circ\dot\gamma=
\dot\mu^i\,\derpar{\phi^i}{v^j}\circ\dot\gamma+
\mu_i(\frac{d}{d t}\Big(\derpar{\Lag}{v^j}\circ\dot\gamma\Big)-
\derpar{\Lag}{q^j}\circ\dot\gamma )\ .
$$ 
These are the {\sl Euler--Lagrange equations of vakononomic mechanics}.

%%%%%%%%%%%%%%%%%%%%%%%%%%%%%%%%%%%%%%%%%
%%%%%%%%%%%%%%%%%%%%%%%%%%%%%%%%%%%%%%%%%%%
%%%%%%%%%%%%%%%%%%%%%%%%%%%%%%%%%%%%%%%%%%%
\subsection{Herglotz variational calculus and contact dynamics}

To study dissipative systems in mechanics using variational techniques, Herglotz generalized Hamilton principle in \cite{He-1930}, introducing a new variable $s$ in the Lagrangian, hence we have 
$\mathfrak{L}\colon \Tan Q \times \mathbb{R} \to \mathbb{R}$, $\mathfrak{L}(q^i,v^i,s)$, and changing the integral functional by a differential one. As we said in the introduction, recently, this approach has given rise to the contact formulation of dissipative mechanics and this kind of Lagrangians are called ``dissipative". In fact this name is not fully accepted and there is another one: ``action dependent Lagrangian" (see \cite{LPAG-2018} and references therein for comments on the subject).

This Section is devoted to give a brief account of Herglotz approach for a Lagrangian of this kind, a dissipative Lagrangian, and obtain the classical dynamical equations for a dissipative system. We also include a comment on the contact formulation.

We will suppose that the Lagrangian is regular in the classical sense with respect to the variables $(q^i,v^i)$, that is the square matrix corresponding to the partial Hessian with respect to the velocities,
$\displaystyle W_{ij} =  \frac{\partial^2\mathfrak{L}}{\partial v^i \partial v^j}$,
is regular at every point.
We simply say that $\mathfrak{L}$ is {\bf regular}; otherwise we say that the Lagrangian is {\bf singular}.

\subsubsection{The variational approach}

The Herglotz principle is a generalization of the Hamilton variational principle in which the Lagrangian depends on the action.

Given a Lagrangian $\mathfrak{L}: \Tan Q \times \mathbb{R} \to \mathbb{R}$, one considers the space of curves with fixed endpoints ${\rm q}_0,{\rm q}_1 \in Q$, which we denote by 
$$
  \Omega({\rm q}_0,{\rm q}_1) = \{\gamma: [0,T] \to Q \mid c(0) ={\rm q}_0, c(T) ={\rm q}_1 \} \ .
$$
We also fix our initial action value $z_0 \in \mathbb{R}$. Given this data, one can compute the action $s_{\gamma}:[0,T] \to \mathbb{R}$ of a curve $\gamma$ by solving the following Cauchy problem:
$$
  \begin{cases}
     \displaystyle \frac{\dd s_{\gamma}}{\dd t} &= \mathfrak{L}(\gamma, \dot{\gamma},  s_{\gamma}),\\
       s_{\gamma}(0) &= z_0 \ .
      \end{cases}
$$
Then we define the \emph{Herglotz action funtional} as 
$$
  \begin{aligned}
      \mathcal{A}: \Omega({\rm q}_0,{\rm q}_1) &\to \mathbb{R},\\
      \gamma &\mapsto s_{\gamma}(T) - s_{\gamma}(0) = \int_0^T \dot{s}_{\gamma} d t = \int_0^T \mathfrak{L}(\gamma, \dot{\gamma},  s_{\gamma})  d t \ .
  \end{aligned}
$$
The critical points of the action $\mathcal{A}$ are the paths $\gamma$ such that $(\gamma, \dot{\gamma}, s_{\gamma})$ is a solution to the \emph{Herglotz equation}~(see~\cite{DeLeon2019} for a proof):
\begin{equation}\label{eq:herglotz}
  \frac{\partial \mathcal{L}}{\partial q^i}-\frac{\d}{\d t}(\frac{\partial \mathfrak{L}}{\partial v^i})=-\frac{\partial \mathfrak{L}}{\partial s}\frac{\partial \mathfrak{L}}{\partial v^i} \ ,
\end{equation}

\subsubsection{The geometric approach}

There is a more geometric approach to the Herglotz equations which shows their link with contact geometry.

A \emph{contact manifold} is a couple $(M, \eta)$ where $M$ is a $(2n+1)$-dimensional manifold and $\eta$ is a one-form such that $\eta \wedge {(d \eta)}^n$ is a volume form. 
We extend the Liouville vector field $\Delta$ and the vertical endomorphism $J$ from $\Tan Q$ to the product manifold $\Tan Q \times \mathbb{R}$. 
Then, given a Lagrangian function $\mathfrak{L}:\Tan Q \times \mathbb{R} \to \mathbb{R}$,
we define the one-form
$$
  \eta_{\mathfrak{L}} = \d s - J^*\d\mathfrak{L}  =\d s - \frac{\partial {\mathfrak{L}}}{\partial v^i}\d q^i,
$$
where $s:\Tan Q \times \mathbb{R} \to \mathbb{R}$ is the projection on the second factor,
and the Lagrangian energy
$$
  E_{\mathfrak{L}} = \Delta(\mathfrak{L}) -\mathfrak{L} = v^i\frac{\partial\mathfrak{L}}{\partial v^i} -\mathfrak{L} \ .
$$

The form $\eta_{\mathfrak{L}}$ is a contact form if, and only if, $\mathfrak{L}$ is regular. In this situation there is a unique vector field $X_{\mathfrak{L}}$ which is the \emph{contact Hamiltonian vector field} of $E_{\mathfrak{L}}$. That is, it satisfies
$$
    \eta_{\mathfrak{L}}(X_{\mathfrak{L}}) = - E_{\mathfrak{L}} 
    \quad , \quad
    \Lie_{X_{\mathfrak{L}}} \eta =  \frac{\partial \mathfrak{L}}{\partial s} \eta \ .
$$
In coordinates, one can see that the integral curves of this vector field are precisely the ones that satisfy the Herglotz equations~\eqref{eq:herglotz}.

If $\mathfrak{L}$ is not regular, the vector field $X_{\mathfrak{L}}$ might not be unique or exist only on a submanifold of $\Tan Q \times \mathbb{R}$. In~\cite{DeLeon2019}, a constraint algorithm was developed in order to deal with this situation.

%%%%%%%%%%%%%%%%%%%%%%%%%%%%%%%%%%%%%%%%%

\section{Herglotz equations as constrained dynamics}

\subsection{Herglotz equations as a nonholonomic ordinary system}

Let $\mathfrak{L}\colon\Tan Q\times\Real\to\Real$, $\mathfrak{L}(q^i,v^i,s)$, be a regular dissipative Lagrangian. 
We try to obtain the dynamics defined by $\mathfrak{L}$ on the nonholonomic submanifold 
of $\Tan(Q\times\Real)$ (endowed with coordinates $(q^i,v^i,s,v_s)$)
defined by the constraint $\phi(q^i,v^i,s,v_s)=v_s-\mathfrak{L}=0$, given by the differential equation taken by Herglotz as defining the generalized variational principle. 
Hence, if $\pi_o\colon\Tan(Q\times\Real)\to\Tan Q\times\Real$ is the canonical projection,
we consider the Lagrangian $L=\pi_o^*\mathfrak{L}\in\Cinfty(\Tan(Q\times\Real))$ as a degenerate Lagrangian since it does not depend on $v_s$, and apply the general theory of degenerate Lagrangians with nonholonomic constraints, the nonholonomic constraint being $\phi(q^i,v^i,s,v_s)=v_s-L=0$.

The canonical Lagrangian forms associated to $L$ are given by
\begin{eqnarray}
\Theta_L&=&\d^\mathrm{v}L=\frac{\partial L}{\partial v^i}\d q^i\ , \nonumber \\
\Omega_L&=&-\d\theta_L=\d q^i\wedge\d(\frac{\partial L}{\partial v^i}) \nonumber \\
&=&
\d q^i\wedge\frac{\partial^2 L}{\partial q^j\partial v^i}\d q^j+
\d q^i\wedge\frac{\partial^2 L}{\partial v^j\partial v^i}\d v^j+
\d q^i\wedge\frac{\partial^2 L}{\partial s\partial v^i}\d s\, .
\label{lagrangforms}
\end{eqnarray}
Denoting by $\Delta$ the Liouville vector field in $\Tan(Q\times\Real)$, that is
$$
\Delta=v_s\frac{\partial}{\partial v_s}+v^i\frac{\partial}{\partial v^i }=\Delta^\Real+\Delta^Q\, ,
$$
 the corresponding Lagrangian energy is
\begin{equation}
\label{lagrangenergy}
E_L=\Delta(L)-L=v_s\frac{\partial L}{\partial v_s}+
v^h\frac{\partial L}{\partial v^h}-L
=v^h\frac{\partial L}{\partial v^h}-L\, ,
\end{equation}
hence
\begin{equation}
\label{diflagrangenergy}
\d E_L= v^h\frac{\partial^2 L}{\partial q^i\partial v^h} \d q^i+
v^h\frac{\partial^2 L}{\partial v^i\partial v^h}\d v^i+
v^h\frac{\partial^2 L}{\partial s\partial v^h}\d s-
\frac{\partial L}{\partial q^i}\d q^i-\frac{\partial L}{\partial s}\d s\, .
\end{equation}

\subsubsection{Formulation with the dynamical vector field}

As a constrained nonholonomic Lagrangian system, the dynamical vector field, $X_L\in\vf(\Tan(Q\times\Real))$, is given by the dynamical equation
\begin{equation}\label{dyneq-nohol}
\inn_{X_L}\Omega_L-\d E_L=\lambda\d^\mathrm{v}\phi\, ,
\end{equation}
where $\lambda:\Real\to\Real$ is the so called Lagrange multiplier.

But $\Omega_L$ is presymplectic, since $\displaystyle\ker\Omega_L=\{\derpar{}{v_s}\}$; hence we need to compute the primary constraints as the compatibility equation
$$
\inn_{\derpar{}{v_s}}(\inn_{X_L}\Omega_L-\d E_L)=
\inn_{\derpar{}{v_s}}(\lambda\d^\mathrm{v}\phi)\, .
$$
Both sides are identically zero because the only term possibly containing $v_s$ is $\d^\mathrm{v}\phi$, but
$$
\d^\mathrm{v}\phi=\frac{\partial \phi}{\partial v_s}\d s+
\frac{\partial \phi}{\partial v^i}\d q^i=\d s-\frac{\partial L}{\partial v^i}\d q^i\, ,
$$
hence
$$
0=\inn_{\derpar{}{v_s}}(\lambda\d^\mathrm{v}\phi)\, ,
$$
that is, we have not any primary constraint, thus the equation for $X_L$ has solution all over the submanifold defined by the constraint, $v_s=L$; that is,  $\{(q^i,v^i,s,v_s)\in\Tan(Q\times\Real)\, ;\ v_s=L(q^i,v^i,s)\}\subset\Tan(Q\times\Real)$.

To calculate the coordinate expression of $X_L$, put
$$
X_L=A^i\frac{\partial }{\partial q^i}+B^i\frac{\partial }{\partial v^i}+
F\frac{\partial }{\partial s}+G\frac{\partial }{\partial v_s}\, ,
$$
and we obtain the following equations for the coefficients of $\d q^i, \d v^i,\d s$ of the dynamical equation(\ref {dyneq-nohol}):
\begin{enumerate}
\item For $\d q^i$,
\begin{equation}\label{b's-free}
A^h\frac{\partial^2 L}{\partial q^i\partial v^h} -A^h\frac{\partial^2 L}{\partial q^h\partial v^i} -B^h\frac{\partial^2 L}{\partial v^h\partial v^i} -F\frac{\partial^2 L}{\partial s\partial v^i} -v^h\frac{\partial^2 L}{\partial q^i\partial v^h} +\frac{\partial L}{\partial q^i}=
-\lambda\frac{\partial L}{\partial v^i}\end{equation}
\item For $\d v^i$,
$$
A^h\frac{\partial^2 L}{\partial v^i\partial v^h} -v^h\frac{\partial^2 L}{\partial v^i\partial v^h}
=0\, ,
$$
which implies that $A^h=v^h$, because $L$ is supposed to be regular.
\item For $\d s$,
$$
A^h\frac{\partial^2 L}{\partial s\partial v^h} -v^h\frac{\partial^2 L}{\partial s\partial v^h}+\frac{\partial L}{\partial s}=\lambda\,.
$$
And, using that $A^h=v^h$, this implies  $\displaystyle\lambda=\derpar{L}{s}$.
\end{enumerate}
The corresponding component of $\d\ v_s$ does not appear.

But, as a consequence of the variational approach,
we have to assume that $X_L$ must be a second order vector field, not only in the coordinates $(q^i,v^i)$ of $\Tan Q$, but also in the dissipation coordinates $(s,v_s)$; therefore we get $\dot{s}=v_s$
and, using the constraint $v_s-L=0$,
we obtain $F=L$. Then, introducing the values of $A^h$ and $\lambda$ in equation (\ref{b's-free}),
we have that
\begin{equation}\label{herg-eq-noholo}
v^h\frac{\partial^2 L}{\partial q^h\partial v^i} -B^h\frac{\partial^2 L}{\partial v^h\partial v^i}+ \frac{\partial^2 L}{\partial s\partial v^i}L-\frac{\partial L}{\partial q^i}=
\frac{\partial L}{\partial s}\frac{\partial L}{\partial v^i}\, .
\end{equation}
from where the coefficients $B^i$ are obtained, because the Lagrangian is regular. Hence, with $G$ still indeterminated, the dynamical vector field is 
$$
X_L=v^i\frac{\partial }{\partial q^i}+B^i\frac{\partial }{\partial v^i}+
L\frac{\partial }{\partial s}+G\frac{\partial }{\partial v_s}\, .
$$
But we need to apply the tangency condition, $\Lie_{X_L}\phi=\Lie_{X_L}(v_s-L)=0$, which gives
$$
G=v^i\frac{\partial L}{\partial q^i}+B^i\frac{\partial L}{\partial v^i}+L\frac{\partial L}{\partial s}\, ,
$$
that is
\begin{equation}\label{nonholodynvf}
X_L=v^i\frac{\partial }{\partial q^i}+B^i\frac{\partial }{\partial v^i}+
L\frac{\partial }{\partial s}+(v^i\frac{\partial L}{\partial q^i}+B^i\frac{\partial L}{\partial v^i}+L\frac{\partial L}{\partial s})\frac{\partial }{\partial v_s}\, .
\end{equation}
Then the dynamical vector field is completely determined
and the integral curves of the vector field $X_L$ are the dynamical trajectories of the system.

There is another way of writing the equations for those trajectories as solutions to a differential equation by means of the Euler-Lagrange operator on the manifold $Q\times\Real$,
$$
\mathcal{E}_L\circ\ddot\gamma=\lambda\d^\mathrm{v}\phi\circ\dot\gamma\,.
$$
In this case the Lagrange multiplier $\lambda$ is the same as above by equation (\ref{dynvf=E-Lop}). In fact we obtain
$$
\frac{\partial L}{\partial q^i}-\frac{\d}{\d t}(\frac{\partial L}{\partial v^i})
=-\lambda\frac{\partial L}{\partial v^i}\nonumber\quad ,\quad
\frac{\partial L}{\partial s}-\frac{\d}{\d t}(0)\nonumber
=\lambda\ ,
$$
that is,
$$
\frac{\partial L}{\partial q^i}-\frac{\d}{\d t}(\frac{\partial L}{\partial v^i})=
-\frac{\partial L}{\partial s}\frac{\partial L}{\partial v^i} \ ,
$$
for a curve $\gamma:\Real\to Q\times\Real$, $(q^i(t),s(t))$, to take the total derivative $\displaystyle\frac{d}{d t}$. 
This is the classical Herglotz equation as we have seen above (see, for example, \cite{GGMRR-2019b} and references therein).

To finish this approach to dissipative systems as nonholonomic constrained systems, we obtain the variation of the energy along the integral curves of the dynamical vector field. From the dynamical equation
$$
\inn_{X_L}\Omega_L-\d E_L=\lambda\d^\mathrm{v}\phi\, ,
$$
using the above result on $\lambda$, we have
$$
\inn_{X_L}\Omega_L=\d E_L +\frac{\partial L}{\partial s}(\d s-\Theta_L)
\, ,
$$
where $\displaystyle\Theta_L=\derpar{L}{v^i}\,\d q^i$.
Then taking the inner contraction with $X_L$,
$$
0=\inn_{X_L}\d E_L +\frac{\partial L}{\partial s}\inn_{X_L}(\d s-\Theta_L)
=\Lie_{X_L}E_L+\frac{\partial L}{\partial s}(L-\Lie_\Delta L)
=\Lie_{X_L} E_L-\frac{\partial L}{\partial s}E_L \ .
$$
Hence
$$
\Lie_{X_L} E_L=\frac{\partial L}{\partial s}E_L \ ,
$$
as in the classical contact formulation (see, for example, \cite{GGMRR-2019b}).

\subsubsection{Formulation with the Euler-Lagrange operator}

The same problem can be addressed from the point of view of the Euler-Lagrange operator in the following way:
The dynamical equation (\ref{dyneq-nohol})
%\begin{equation}
%\inn_{X_L}\Omega_L-\d E_L=\lambda\d^\mathrm{v}\phi\, ,
%\end{equation}
is now written as $\mathcal{E}_L=\lambda\d^\mathrm{v}\phi$, and 
by equation (\ref{E-L-product}), it can be decomposed in the form
$$
\mathcal{E}_L=\mathcal{E}_L^{Q}+\mathcal{E}_L^{\Real}=
(\frac{\partial L}{\partial q^i}-\frac{\d}{\d t}(\frac{\partial L}{\partial v^i}))\d q^i
+(\frac{\partial L}{\partial s}-\frac{\d}{\d t}(\frac{\partial L}{\partial v_s}))\d s=
\mathcal{E}_L^{Q}
+\frac{\partial L}{\partial s}\d s\ .
$$
Furthermore, from (\ref{d^v-product}), we have
$$
\d^\mathrm{v}\phi=\d_Q^\mathrm{v}\phi+\d_\Real^\mathrm{v}\phi=
\d_Q^\mathrm{v}v_s-\d_Q^\mathrm{v}L+\d_\Real^\mathrm{v}v_s+\d_\Real^\mathrm{v}L
=-\frac{\partial L}{\partial v^i}\d q^i+\d s\ .
$$
Hence we have
$$
\mathcal{E}_L^{Q}+\frac{\partial L}{\partial s}\d s=-\lambda\frac{\partial L}{\partial v^i}\d q^i
+\lambda\d s\, ;
$$
which gives
$$
\mathcal{E}_L^{Q}=-\lambda\frac{\partial L}{\partial v^i}\d q^i\quad ,\quad
\frac{\partial L}{\partial s}=\lambda\, ;
$$
and the dynamical equation is
$$
\mathcal{E}_L^{Q}=-\frac{\partial L}{\partial s}\frac{\partial L}{\partial v^i}\d q^i\,,
$$
which is the classical Herglotz equation for dissipative Lagrangian systems. This equation can be also write as
$$
\mathcal{E}_L^{Q}=-\frac{\partial L}{\partial s}\d_Q^\mathrm{v}L\,.
$$ 

\subsection{Summary of Section 3.1}

In this section, we have studied  the dissipative Lagrangian system defined by the regular dissipative Lagrangian 
$\mathfrak{L}\colon\Tan Q\times\Real\to\Real$, $\mathfrak{L}(q^i,v^i,s)$ and two associated dynamics. 

First, in Section 2.5, we have obtained the associated dynamics as a contact Hamiltonian system defined by $(\Tan Q\times\Real,\eta_{\mathfrak{L}},E_{\mathfrak{L}})$. This corresponds to the geometric version of the Herglotz principle for the Lagrangian $\mathfrak{L}$.

Second, in this Section 3.1, we have considered the constrained Lagrangian system defined by 
 $(\Tan(Q\times\Real),L=\pi_o^*\mathfrak{L})$, with the nonholonomic constraint $\phi(q^i,v^i,s,v_s)=v_s-L=0$, and have obtained the corresponding dynamics.
 
 And we have proven that.
 
 \begin{teor}
 Both systems, the contact one $(\Tan Q\times\Real,\eta_{\mathfrak{L}},E_{\mathfrak{L}})$, and the nonholonomic Lagrangian one $(\Tan(Q\times\Real),L=\pi_o^*\mathfrak{L})$, with the nonholonomic constraint $\phi(q^i,v^i,s,v_s)=v_s-L=0$, have the same dynamics.
 The differential equations of the solution trajectories are the same.
 \end{teor}
 
 As a consequence, the appropriate dynamics of the dissipative Lagrangian systems can be studied under the approach of symplectic, or presymplectic, nonholonomic dynamics instead of the contact approach.
%%%%%%%%%%%%%%%%%%%%%%%%%%%%%%%%%%%%%%%%%

\subsection{Herglotz equations as a vakonomic ordinary system}

As in the previous section, we consider a dissipative regular Lagrangian 
$\mathfrak{L}\colon\Tan Q\times\Real\to\Real$, $\mathfrak{L}(q^i,v^i,s)$, 
constrained now by a vakonomic constraint in
$\Tan(Q\times\Real)$ given by $\psi(q^i,v^i,s,v_s)=v_s-L=0$. 
We have a singular Lagrangian $L$ defined on $\Tan(Q\times\Real)$ 
with a non integrable vakonomic constraint
(as above, we have denoted $L=\pi_o^*\mathfrak{L}\in\Cinfty(\Tan(Q\times\Real))$ and the vakonomic constraint being $\psi(q^i,v^i,s,v_s)=v_s-L=0$).

\subsubsection{Formulation with the dynamical vector field}

As recalled in Section 2.4, the vakonomic dynamical vector field $Y_L\in\vf(\Tan(Q\times\Real))$ is the solution to the equation
$$
\inn_{Y_L}\Omega_L-\d E_L=\mathcal{E}_{\mu\psi}=\dot{\mu}\d^\mathrm{v}\psi+\mu\mathcal{E}_\psi\, ,
$$
where $\mu:\Real\to\Real$ is the Lagrange multiplier and $\mathcal{E}_\psi$ is the Euler--Lagrange operator, in the manifold $Q\times\Real$, for the function $\psi$ which in coordinates gives, for $\psi=v_s-L$,
\begin{eqnarray*}
\mathcal{E}_\psi&= &(-\frac{\partial L}{\partial s}+\frac{\d}{\d t}(1))\d s-(\frac{\partial L}{\partial q^i}-\frac{\d}{\d t}(\frac{\partial L}{\partial v^i}))\d q^i \\
&=&
-\frac{\partial L}{\partial s}\d s-(\frac{\partial L}{\partial q^i}-\frac{\d}{\d t}(\frac{\partial L}{\partial v^i}))\d q^i \ , \\
\d^\mathrm{v}\psi&=&\d_Q^\mathrm{v}\psi+\d_Q^\mathrm{v}\psi =\d s-\frac{\partial L}{\partial v^i}\d q^i
\ .
\end{eqnarray*} 
To compute the local expression of $Y_L$, we write
$$
Y_L=A^i\frac{\partial }{\partial q^i}+B^i\frac{\partial }{\partial v^i}+
F\frac{\partial }{\partial s}+G\frac{\partial }{\partial v_s}\, ,
$$
and obtain the following equations:
\begin{enumerate}
\item For $\d q^i$, 
\begin{eqnarray}\label{b's-vak}
A^h\frac{\partial^2 L}{\partial q^i\partial v^h} &-&A^h\frac{\partial^2 L}{\partial q^h\partial v^i} -B^h\frac{\partial^2 L}{\partial v^h\partial v^i}-F\frac{\partial^2 L}{\partial s\partial v^i} -v^h\frac{\partial^2 L}{\partial q^i\partial v^h} +\frac{\partial L}{\partial q^i}\nonumber
\\ &=&
-\dot{\mu}\frac{\partial L}{\partial v^i}
-\mu\frac{\partial L}{\partial q^i}+
\mu(\frac{\partial^2 L}{\partial q^j\partial v^i}v^j+  \frac{\partial^2 L}{\partial v^j\partial v^i}B^j+ \frac{\partial^2 L}{\partial s\partial v^i}v_s)\, .
\end{eqnarray}
\item For $\d v^i$,
$$
A^h\frac{\partial^2 L}{\partial v^i\partial v^h} -v^h\frac{\partial^2 L}{\partial v^i\partial v^h}
=0\, ,
$$
which implies that $A^h=v^h$ because $L$ is supposed to be regular.
\item For $\d s$,
$$
A^h\frac{\partial^2 L}{\partial s\partial v^h} -v^h\frac{\partial^2 L}{\partial s\partial v^h}+\frac{\partial L}{\partial s}=\dot{\mu} -\mu\frac{\partial L}{\partial s}  \,.
$$
And, as $A^h=v^h$ by item 2, the multiplier $\mu$ satisfies the differential equation
\begin{equation}\label{dotmu}
\frac{\partial L}{\partial s}=\dot{\mu} -\mu\frac{\partial L}{\partial s} \, ;
\end{equation}
that is,
$$
\dot{\mu} =(1+\mu)\frac{\partial L}{\partial s}\, .
$$
\end{enumerate}
The component that corresponds to $\d v_s$ does not appear.

 As above, assuming that $Y_L$ must be a second order vector field also in the dissipation coordinates $(s,v_s)$, we get $\dot{s}=v_s$, and
the constraint $\psi$ gives $F=v_s=L$. Then, by substitution of $A^h=v^h$ and equation (\ref{dotmu}) into (\ref{b's-vak}), we obtain
\begin{eqnarray*}
&-&v^h\frac{\partial^2 L}{\partial q^h\partial v^i} -B^h\frac{\partial^2 L}{\partial v^h\partial v^i}-v_s\frac{\partial^2 L}{\partial s\partial v^i} 
+\frac{\partial L}{\partial q^i}\\
&=&
-(1+\mu)\frac{\partial L}{\partial s}\frac{\partial L}{\partial v^i}
-\mu\frac{\partial L}{\partial q^i}+
\mu(\frac{\partial^2 L}{\partial q^j\partial v^i}v^j+  \frac{\partial^2 L}{\partial v^j\partial v^i}B^j+ \frac{\partial^2 L}{\partial s\partial v^i}v_s) \ ;
\end{eqnarray*}
that is, with $F=v_s=L$,
$$
(1+\mu)[-v^h\frac{\partial^2 L}{\partial q^h\partial v^i} -B^h\frac{\partial^2 L}{\partial v^h\partial v^i}- \frac{\partial^2 L}{\partial s\partial v^i}L+\frac{\partial L}{\partial q^i}]=
-(1+\mu)\frac{\partial L}{\partial s}\frac{\partial L}{\partial v^i}\, .
$$
Remark that $\mu\neq -1$ because $\mu=-1$ implies that the dynamical equation is 
$\inn_{Y_L}\Omega_L-\d E_L=\mathcal{E}_{-\psi}$ which is an equivalence because $-\psi=L-v_s$ and $v_s$ has an Euler-Lagrange operator identically zero. Hence $1+\mu\neq 0$, and we have
\begin{equation}\label{herg-eq}
v^h\frac{\partial^2 L}{\partial q^h\partial v^i} -B^h\frac{\partial^2 L}{\partial v^h\partial v^i}+ \frac{\partial^2 L}{\partial s\partial v^i}L-\frac{\partial L}{\partial q^i}=
\frac{\partial L}{\partial s}\frac{\partial L}{\partial v^i}\, .
\end{equation}
from where we can obtain the coefficients $B^i$ 
because the Lagrangian $L$ is regular. 
To obtain the component $G$ we use the tangency condition, as in the nonholonomic case, and obtain the same solution.

As equation (\ref{herg-eq}) coincides with equation (\ref{herg-eq-noholo}), we have obtained the Herglotz dissipative dynamics as a classical Lagrangian system with a vakonomic constraint given by the original differential equation used by Herglotz to generalize the classical Lagrangian variational mechanics.

Then we have determined completely the dynamical vector field $Y_L$ and it is the same we obtained in (\ref{nonholodynvf}), the corresponding to the nonholonomic case: both give the Herglotz dissipative dynamics.

Observe that equation (\ref{herg-eq}) can be writen as
$$
\frac{\partial L}{\partial q^i}-\frac{\d}{\d t}(\frac{\partial L}{\partial v^i})=-\frac{\partial L}{\partial s}\frac{\partial L}{\partial v^i} \ ,
$$
which is, once again, the well known coordinate expression of the Herglotz equations for the curves $\gamma:\Real\to Q\times\Real$, $\gamma(t)=(q^i(t),s(t))$, for Lagrangian dissipative systems given by $L(q^i,v^i,s)$.

As the obtained dynamical vector fields are $X_L=Y_L$, then the variation of the Lagrangian energy along the trajectories has the same expression.
$$
\Lie_{Y_L} E_L=\frac{\partial L}{\partial s}E_L \ .
$$

\subsubsection{Formulation with the Euler-Lagrange operator}

In this approach the dynamical equation for the system is written, see equation (\ref{vELo}), as
$$
\mathcal{E}_L=\mathcal{E}_{\mu\psi}=\dot{\mu}\d^\mathrm{v}\psi+\mu\mathcal{E}_\psi\ .
$$
But by equation (\ref{E-L-product}), the last equation
can be decomposed in the form
$$
\mathcal{E}_L=\mathcal{E}_L^{Q}+\mathcal{E}_L^{\Real}=
(\frac{\partial L}{\partial q^i}-\frac{\d}{\d t}(\frac{\partial L}{\partial v^i}))\d q^i
+(\frac{\partial L}{\partial s}-\frac{\d}{\d t}(\frac{\partial L}{\partial \dot{s}}))\d s=
\mathcal{E}_L^{Q}+\frac{\partial L}{\partial s}\d s\, ,
$$
and
\begin{eqnarray*}
\mathcal{E}_\psi&=&\mathcal{E}_\psi^{Q}+\mathcal{E}_\psi^{\Real}=
\mathcal{E}_{\dot{s}}^{Q}+\mathcal{E}_{\dot{s}}^{\Real}-\mathcal{E}_L^{Q}-\mathcal{E}_L^{\Real}
= - \mathcal{E}_L^{Q}-\frac{\partial L}{\partial s}\d s  \ .
\end{eqnarray*}
Hence $\mathcal{E}_\psi=-\mathcal{E}_L$.

On the other side, from equation (\ref{d^v-product}), we have
$$
\d^\mathrm{v}\psi=\d_Q^\mathrm{v}\psi+\d_\Real^\mathrm{v}\psi=
\d_Q^\mathrm{v}v_s-\d_Q^\mathrm{v}L+\d_\Real^\mathrm{v}v_s+\d_\Real^\mathrm{v}L
=-\frac{\partial L}{\partial v^i}\d q^i+\d s\,.
$$
Then we have
$$
\mathcal{E}_L^{Q}+\frac{\partial L}{\partial s}\d s=-\dot{\mu}\frac{\partial L}{\partial v^i}\d q^i
+\dot{\mu}\d s
-\mu\mathcal{E}_L^{Q}-\mu\frac{\partial L}{\partial s}\d s \, ;
$$
that is
$$
\mathcal{E}_L^{Q}=-\dot{\mu}\frac{\partial L}{\partial v^i}\d q^i-\mu\mathcal{E}_L^{Q}
\quad ,\quad
\frac{\partial L}{\partial s}=\dot{\mu}-\mu\frac{\partial L}{\partial s}
\,.
$$
If we substitute the last equation into the previous one and reorder terms, we obtain:
\begin{equation}\label{A-mu+1}
(\mu+1)\mathcal{E}_L^{Q}=-(\mu+1)\frac{\partial L}{\partial s}\frac{\partial L}{\partial v^i}\d q^i\,.
\end{equation}
But $\mu$ can not be equal to $-1$ because, from the above calculation, we know that 
$\mathcal{E}_\psi=-\mathcal{E}_L$, hence if $\mu=-1$ then the dynamical equation is no more that $\mathcal{E}_L=\mathcal{E}_L$,
which gives no specific dynamics associated. Hence, dividing by $1+\mu$, we have
$$
\mathcal{E}_L^{Q}=-\frac{\partial L}{\partial s}\frac{\partial L}{\partial v^i}\d q^i,
$$
which is the classical Herglotz equation for dissipative Lagrangian systems. In a more intrinsic way, we can write
$$
\mathcal{E}_L^{Q}=-\frac{\partial L}{\partial s}\d_Q^\mathrm{v}L\,.
$$

\subsection{Summary of Section 3.3}

In a similar way as in Section 3.1, here we continue the study of the dissipative Lagrangian system defined by the regular dissipative Lagrangian 
$\mathfrak{L}\colon\Tan Q\times\Real\to\Real$, $\mathfrak{L}(q^i,v^i,s)$ and two associated dynamics. 

First, in Section 2.5, as a contact Hamiltonian system defined by $(\Tan Q\times\Real,\eta_{\mathfrak{L}},E_{\mathfrak{L}})$. This corresponds to the geometric version of the Herglotz principle for the Lagrangian $\mathfrak{L}$.

Second, in this Section 3.2, we have considered the constrained Lagrangian system defined by 
 $(\Tan(Q\times\Real),L=\pi_o^*\mathfrak{L})$, with the constraint $\psi(q^i,v^i,s,v_s)=v_s-L=0$ considered as a vakonomic system, and have obtained the corresponding dynamics. Observe that this is another different approach to the one in Section 3.1. Here we consider the same non integrable constraint but the system is considered as a vakonomic one.
 
 The result we have proven is:
 
\begin{teor}
 Both systems, the contact one $(\Tan Q\times\Real,\eta_{\mathfrak{L}},E_{\mathfrak{L}})$, and the vakonomic Lagrangian one $(\Tan(Q\times\Real),L=\pi_o^*\mathfrak{L})$, with the constraint $\psi(q^i,v^i,s,v_s)=v_s-L=0$, have the same dynamics.
 The differential equations of the solution trajectories are the same.
 \end{teor}
 
 As a consequence, the dissipative Lagrangian systems can be studied under the approach of symplectic, or presymplectic, vakonomic dynamics instead of the contact approach. The equivalence of both the nonholonomic and the vakonomic approaches is due to the specific expressions of the Lagrangian, not depending on $v_s$, and the constraint. 
%%%%%%%%%%%%%%%%%%%%%%%%%%%%%%%%%%%%%%%%%

\section{Nonholonomic constrained dissipative Lagrangians}

Let $\mathfrak{L}:\Tan(Q\times\Real)\to\Real$, $\mathfrak{L}(q^i,v^i,s)$, be a regular dissipative Lagrangian and $L=\pi_o^*\mathfrak{L}\in\Cinfty(\Tan(Q\times\Real))$, as in the previous sections. 
In this Section we analyze the dynamics determined by $L$ constrained to  move, for the dynamical variables $(q^i,v^i)$, in a submanifold $N\subset\Tan Q$
defined by a family of constraints $\phi^\alpha:\Tan Q\to\Real$, $\alpha=1,\ldots, h$,  
independent with respect to the velocities $(v^1,\ldots,v^n)$; that is,
$$
\mathrm{rank}(\frac{\partial(\phi^1,\ldots,\phi^h)}{\partial(v^1,\ldots,v^n)})=h\,.
$$
We understand that the manifold $N$ is included in $\Tan(Q\times\Real)$. We consider $N$ as a nonholonomic constraint submanifold in this paragraph.

As a dissipative system we consider the dynamics is also subjected to the dissipative constraint given by $\phi=v_s-L=0$ as another nonholonomic constraint. Hence the constraint submanifold $C\subset N\subset\Tan(Q\times\Real)$ is defined by the family of constraints $\{\phi,\phi^\alpha\}$. Observe that they are independent with respect to the velocities $(v_s,v^1,\ldots,v^n)$; that is,
$$
\mathrm{rank}(\frac{\partial(\phi,\phi^1,\ldots,\phi^h)}{\partial(v_s,v^1,\ldots,v^n)})=h+1\,.
$$

\noindent{\bf Comment on the above hypotheses}: We know that the dynamical variables are $(q^i,v^i)$. The variable $s$ is introduced to give account of the dissipative process and the Herglotz approach. As the constraints must be in the dynamical variables, it is natural to assume that the functions $\phi^\alpha$ depend only on $(q^i,v^i)$ and not on $(s,v_s)$. Thus we have a submanifold $N\subset\Tan Q$ and we consider it as the phase space where we study the motion. Nevertheless, in Section \ref{4-3} we will comment the general situation.

Summarizing this discussion, we consider that $N\subset\Tan(Q\times\Real)$ is a nonholonomic constraint submanifold and the system under study, defined by the Lagrangian $L$ and the constraints $\{\phi,\phi^\alpha\}$, as a nonholonomic system. Observe that we give the same name to the constraints $\phi, \phi^\alpha$ and their corresponding backup from $\Tan Q\times\Real$ or$\Tan Q$ to $\Tan(Q\times\Real)$ respectively.

\subsection{Formulation with the dynamical vector field}

We are working with a constrained nonholonomic Lagrangian system which is defined by $(Q\times\Real,L,\phi,\phi^\alpha)$, thus the dynamical vector field, $X_L\in\vf(\Tan(Q\times\Real))$, is given by the equation
\begin{equation}\label{nonholodyneq}
\inn_{X_L}\Omega_L-\d E_L=\lambda\d^\mathrm{v}\phi+
\lambda_\alpha\d^\mathrm{v}\phi^\alpha\, ,
\end{equation}
where $\lambda,\lambda_\alpha:\Real\to\Real$ are the Lagrange multipliers.
The canonical Lagrangian forms and the energy associated to $L$ are the same as above in equations (\ref{lagrangforms}, \ref{lagrangenergy}, \ref{diflagrangenergy}). 

As above, $\Omega_L$ is presymplectic, since $\displaystyle\ker\Omega_L=\{\derpar{}{v_s}\}$, hence we need to compute the primary constraints as the compatibility equation
$$
\inn_{\derpar{}{v_s}}(\inn_{X_L}\Omega_L-\d E_L)=
\inn_{\derpar{}{v_s}}(\lambda\d^\mathrm{v}\phi+
\lambda_\alpha\d^\mathrm{v}\phi^\alpha)\, .
$$
Both sides are identically zero, hence we have not any primary constraint as above.

Once again, to calculate the coordinate expression of $X_L$, put
$$
X_L=A^i\frac{\partial }{\partial q^i}+B^i\frac{\partial }{\partial v^i}+
F\frac{\partial }{\partial s}+G\frac{\partial }{\partial v_s}\, ,
$$
and, in this case, we obtain the following equations:
\begin{enumerate}
\item For $\d q^i$,
\begin{eqnarray}
A^h\frac{\partial^2 L}{\partial q^i\partial v^h} -A^h\frac{\partial^2 L}{\partial q^h\partial v^i} -B^h\frac{\partial^2 L}{\partial v^h\partial v^i} &-&F\frac{\partial^2 L}{\partial s\partial v^i} -v^h\frac{\partial^2 L}{\partial q^i\partial v^h} +\frac{\partial L}{\partial q^i}\nonumber\\
&=&
-\lambda\frac{\partial L}{\partial v^i}+\lambda_\alpha \frac{\partial \phi^\alpha}{\partial v^i} \ .
\label{b's}
\end{eqnarray}
\item For $\d v^i$,
$$
A^h\frac{\partial^2 L}{\partial v^i\partial v^h} -v^h\frac{\partial^2 L}{\partial v^i\partial v^h}
=0\, ,
$$
which implies that $A^h=v^h$ because $L$ is supposed to be regular.
\item For $\d s$,
$$
A^h\frac{\partial^2 L}{\partial s\partial v^h} -v^h\frac{\partial^2 L}{\partial s\partial v^h}+\frac{\partial L}{\partial s}=\lambda\,.
$$
And, being $v^h=A^h$ by item 2, this implies  $\displaystyle\lambda=\frac{\partial L}{\partial s}$.
\end{enumerate}
And $\d\ v_s$ does not appear.

Assuming again that $X_L$ must be a second order vector field also in the dissipation coordinates $(s,v_s)$, we get $\dot{s}=v_s$
and, using the constraint $v_s-L=0$,
we obtain $F=L$. Then, introducing the values of $A^h$ and $\lambda$ in equation (\ref{b's}),
we have that
$$
X_L=v^i\frac{\partial }{\partial q^i}+B^i\frac{\partial }{\partial v^i}+
L\frac{\partial }{\partial s}+G\frac{\partial }{\partial \dot{s}}\, ,
$$
where, using that the Lagrangian is regular, the coefficients $B^i$ are obtained form equation (\ref{b's}), which by substitution of $A^h=v^h$ reads
$$
 -A^h\frac{\partial^2 L}{\partial q^h\partial v^i} -B^h\frac{\partial^2 L}{\partial v^h\partial v^i} -L\frac{\partial^2 L}{\partial s\partial v^i}+\frac{\partial L}{\partial q^i}=
-\frac{\partial L}{\partial s}\frac{\partial L}{\partial v^i}+\lambda_\alpha \frac{\partial \phi^\alpha}{\partial v^i} \ ,
$$
and $G$ is still indeterminated.

Applying the tangency condition, $\Lie_{X_L}\phi=\Lie_{X_L}(v_s-L)=0$, we obtain
$$
G=v^i\frac{\partial L}{\partial q^i}+B^i\frac{\partial L}{\partial v^i}+L\frac{\partial L}{\partial s}\, ;
$$
that is,
$$
X_L=v^i\frac{\partial }{\partial q^i}+B^i\frac{\partial }{\partial v^i}+
L\frac{\partial }{\partial s}+(v^i\frac{\partial L}{\partial q^i}+B^i\frac{\partial L}{\partial v^i}+L\frac{\partial L}{\partial s})\frac{\partial }{\partial v_s}\, .
$$
Observe that we need to consider the constraints $\phi^\alpha$ as additional equations in order to determine the Lagrange multipliers $\lambda_\alpha$ and, consequently, to obtain the coefficients $B^i$. Then the dynamical vector field is completely determined.

The dynamical trajectories of the system are the integral curves of the dynamical vector field $X_L$ and we can give them as solutions to a differential equation written in terms of the Euler-Lagrange operator. The dynamical equation (\ref{nonholodyneq}) is also given by 
$$
\mathcal{E}_L=\lambda\d^\mathrm{v}\phi+
\lambda_\alpha\d^\mathrm{v}\phi^\alpha\, ,
$$
which in natural coordinates in $\Tan(Q\times \Real)$ reads
$$
\frac{\partial L}{\partial q^i}-\frac{\d}{\d t}\left(\frac{\partial L}{\partial v^i}\right)=-\frac{\partial L}{\partial s}\frac{\partial L}{\partial v^i}
+\lambda_\alpha \frac{\partial \phi^\alpha}{\partial v^i}\, ,
$$
system with $n$ differential equations on the dynamical trajectories $\gamma:\Real\to Q\times\Real$, where the new unknowns $\lambda_\alpha$ are obtained adding to this system the new equations given by the nonholonomic constraints $\phi^\alpha=0$. 

Once again, we can write this last equation in a more intrinsic way as
$$
\mathcal{E}_L^Q=-\frac{\partial L}{\partial s}\d_Q^\mathrm{v}L+
\lambda_\alpha\d_Q^\mathrm{v}\phi^\alpha\, .
$$
Which for the dynamical trajectories $\gamma:\Real\to\Real$ reads
$$
\mathcal{E}_L^Q\circ\ddot\gamma=-\frac{\partial L}{\partial s}\d_Q^\mathrm{v}L\circ\dot\gamma+
\lambda_\alpha\d_Q^\mathrm{v}\phi^\alpha\circ\dot\gamma\, .
$$

With respect to the energy, its variation is given by
$$
\Lie_{X_L}E_L=\frac{\partial L}{\partial s}E_L+\lambda_\alpha\inn_{X_L}\d^\mathrm{v}\phi^\alpha=\frac{\partial L}{\partial s}E_L
+\lambda_\alpha v^i\frac{\partial \phi^\alpha}{\partial v^i}
\frac{\partial L}{\partial s}E_L=
\frac{\partial L}{\partial s}E_L+\lambda_\alpha \Lie_{(\Delta^Q)}\phi^\alpha\ ,
$$
and the last term corresponds to the nonholonomic constraints, being $\Delta^Q$ the extension to $Q\times\Real$ of the Liouville vector field on the manifold $Q$.

\subsection{Formulation with the Euler-Lagrange operator}

As we said above another way to write the equation for the dynamical trajectories of the system is 
$$
\mathcal{E}_L=\lambda\d^\mathrm{v}\phi+
\lambda_\alpha\d^\mathrm{v}\phi^\alpha\, .
$$
With the same decomposition for the operators $\mathcal{E}_L$ and $\d^\mathrm{v}$, this equations is written as
$$
\mathcal{E}_L^Q+\frac{\partial L}{\partial s}\d s=-\lambda\frac{\partial L}{\partial v^i}\d q^i+
\lambda\d s+\lambda_\alpha\frac{\partial \phi^\alpha}{\partial v^i}\d q^i\,.
$$
Identifying components in $\d s$ and $\d q^i$ in both sides of the equation we have that $\displaystyle\lambda=\frac{\partial L}{\partial s}$ and
\begin{equation}\label{eq:nh_eom}
    \mathcal{E}_L^Q=-\frac{\partial L}{\partial s}\d_Q^\mathrm{v}L+
\lambda_\alpha\d_Q^\mathrm{v}\phi^\alpha\, ,
\end{equation}
as above.

\subsection{Summary of Section 4}

We have studied the dynamics of a dissipative Lagrangian $\mathfrak{L}\colon\Tan Q\times\Real\to\Real$, $\mathfrak{L}(q^i,v^i,s)$ as a dissipative system and constrained to a submanifold $N\subset\Tan Q$
defined by a family of constraints $\phi^\alpha:\Tan Q\to\Real$, $\alpha=1,\ldots, h$,  
independent with respect to the velocities $(v^1,\ldots,v^n)$, as a nonholonomic system given by the Lagrangian system $(\Tan(Q\times\Real),L=\pi_o^*\mathfrak{L})$ constrained by the family 
$(\phi(q^i,v^i,s,v_s)=v_s-L=0, \phi^\alpha=0)$ and we have obtained the following

\begin{teor}
The dissipative dynamics corresponding to the dissipative Lagrangian system $(\Tan(Q\times\Real),L=\pi_o^*\mathfrak{L})$ constrained by the family $\phi^\alpha:\Tan Q\to\Real$, $\alpha=1,\ldots, h$, of nonholonomic constraints, is given by
$$
\mathcal{E}_L=\lambda\d^\mathrm{v}\phi+
\lambda_\alpha\d^\mathrm{v}\phi^\alpha\, ,
$$
where $
\mathcal{E}_L$   is the Euler--Lagrange operator.

Which for the trajectories of the system, $\gamma:\Real\to Q\times\Real$, gives:
$$\mathcal{E}_L^Q\circ\ddot\gamma=-\frac{\partial L}{\partial s}\d_Q^\mathrm{v}L\circ\dot\gamma+
\lambda_\alpha\d_Q^\mathrm{v}\phi^\alpha\circ\dot\gamma\, ,
$$
and in natural coordinates in $\Tan(Q\times \Real)$ reads
$$
\frac{\partial L}{\partial q^i}-\frac{\d}{\d t}\left(\frac{\partial L}{\partial v^i}\right)=-\frac{\partial L}{\partial s}\frac{\partial L}{\partial v^i}
+\lambda_\alpha \frac{\partial \phi^\alpha}{\partial v^i}\, ,
$$
system with $n$ differential equations on the dynamical trajectories, where the new unknowns $\lambda_\alpha$ are obtained adding to this system the new equations given by the nonholonomic constraints $\phi^\alpha=0$. 

\end{teor}

Observe that the equations are the same as for a standard nonholonomic systems with constraints $\phi^\alpha$ plus the dissipation term given by
$$
-\frac{\partial L}{\partial s}\frac{\partial L}{\partial v^i}\, ,
$$
which is the usual dissipation term for the dynamics of the dissipative systems obtained by the Herglotz principle.

As an easy example, if we take the standard vertical rolling disk without sliding, but with dissipation term $\delta s$, given by the Lagrangian $\mathcal{L}$ and constraints $\phi_1,\phi_2$
$$\mathcal{L}=\frac{1}{2}(\dot{x}^2+\dot{y}^2+\dot{\theta}^2+\dot{\varphi}^2)+\delta s,\quad  \phi_1=\dot{x}-\dot{\theta}\cos \varphi,\quad 
\phi_2=\dot{y}-\dot{\theta}\sin \varphi\, ,
$$
with $\delta$ a positive real number. The dynamical equations are given by
$$
-\ddot{x}=-\delta \dot{x}+\lambda_1,\quad -\ddot{y}=-\delta \dot{y}+\lambda_2,\quad  -\ddot{\theta}=-\delta \dot{\theta}-\lambda_1 \cos\varphi-\lambda_2\sin\varphi,\quad
-\ddot{\varphi}=-\delta \dot{\varphi}
$$
as can be easily calculated from the above expressions. The constraints must be added to the equations to obtain the trajectories and the Lagrange multipliers because we have a set of four ordinary differential equations with six unknowns.

\subsection{The general case for constraints}
\label{4-3}

If the constraints $\phi^\alpha$ depend on the dissipation variable $s$, the formulation and the results are the same because
$$
\d^\mathrm{v}\phi^\alpha=\d_Q^\mathrm{v}\phi^\alpha \ ,
$$
if $\phi^\alpha=\phi^\alpha(q^i,v^i,s)$ instead of being only functions of $(q^i,v^i)$.

\subsection{Example of a nonholonomic system: Chaplygin's sleigh with dissipation}

We reproduce an example of a nonholonomic system; the Chaplygin's sleigh with dissipation, which can be found in~\cite{LJL-2021}.

The configuration space is $Q= \mathbb{R}^2 \times S^1$, with coordinates $(x,y)$, describing the position of a blade and an angle $\theta$, which describes its rotation. We  added an extra term $\gamma z$, which accounts for friction with the medium with a coefficient $\gamma$.  The Lagrangian is
\begin{equation}
\begin{aligned}
    L=&
    \frac{1}{2} \, {({({\alpha} \cos({\theta}) - {\beta} \sin({\theta}))} {\dot{\theta}} + {\dot{y}})}^{2} + \\ & \frac{1}{2} \, {({({\beta} \cos({\theta}) + {\alpha} \sin({\theta}))} {\dot{\theta}} - {\dot{x}})}^{2} + {\dot{\theta}}^{2} + {\gamma} s
\end{aligned}
\end{equation}
 $(\alpha,\beta)$ is the position of the sleight center of gravity The units are normalized so that the mass and the radius of inertia of the sleight is $1$.

The constraint due to the Lagrangian is
\begin{equation}
\phi = v_s  - L
\end{equation}

The sleight is forced to move in the direction of the blade. This is expressed by the following constraint
\begin{equation}
    \phi^1 = \dot{x} \sin(\theta) - \dot{y} \cos(\theta).
\end{equation}

The equations of motion are given by~\eqref{eq:nh_eom}, where the Euler-Lagrange operator is given by
\begin{equation}
\begin{gathered}
    \mathcal{E}^Q_L = 
    \ddot{x}( -{\beta} \cos({\theta}) - {\alpha} \sin({\theta}) + 1 ) \mathrm{d} x + \ddot{y}( {\alpha} \cos({\theta}) - {\beta} \sin({\theta}) + 1 ) \mathrm{d} y \\ + 
    ( {({({\beta} \cos({\theta}) + {\alpha} \sin({\theta}))} {\dot{\theta}}^{2} - {({\alpha} \cos({\theta}) - {\beta} \sin({\theta}))} {\dot{\theta}})} {\dot{x}}\\ - {({({\alpha} \cos({\theta}) - {\beta} \sin({\theta}))} {\dot{\theta}}^{2} + {({\beta} \cos({\theta}) + {\alpha} \sin({\theta}))} {\dot{\theta}})} {\dot{y}}   \\ +  \ddot{\theta}( {\alpha}^{2} + {\beta}^{2} + {({\alpha} - {\beta})} \cos({\theta}) - {({\alpha} + {\beta})} \sin({\theta}) + 2 ) ) \mathrm{d} {\theta}
\end{gathered}
\end{equation}
We also have that
\begin{equation}
    \frac{\partial L}{\partial s} = \gamma
\end{equation}
\begin{equation}
\begin{gathered}
    \dd^v_Q L = \left( -{\left({\beta} \cos\left({\theta}\right) + {\alpha} \sin\left({\theta}\right)\right)} {\dot{\theta}} + {\dot{x}} \right) \mathrm{d} x + \left( {\left({\alpha} \cos\left({\theta}\right) - {\beta} \sin\left({\theta}\right)\right)} {\dot{\theta}} + {\dot{y}} \right) \mathrm{d} y \\ + \left( {\left({\alpha}^{2} + {\beta}^{2} + 2\right)} {\dot{\theta}} - {\left({\beta} \cos\left({\theta}\right) + {\alpha} \sin\left({\theta}\right)\right)} {\dot{x}} + {\left({\alpha} \cos\left({\theta}\right) - {\beta} \sin\left({\theta}\right)\right)} {\dot{y}} \right) \mathrm{d} {\theta}
    \end{gathered}
\end{equation}
\begin{equation}
    \dd^v_Q \theta^1 = \sin\left({\theta}\right) \mathrm{d} x -\cos\left({\theta}\right) \mathrm{d} y
\end{equation}
%%%%%%%%%%%%%%%%%%%%%%%%%%%%%%%%%%%%%%%%%%%

\section{Dissipative Lagrangian systems with vakonomic constraints}

In this case we consider a dissipative regular Lagrangian $\mathfrak{L}(q^i,v^i,s)\in\Cinfty(\Tan Q\times\Real)$ and we analize the dynamics corresponding to $L=\pi_o^*\mathfrak{L}\in\Cinfty(\Tan(Q\times\Real))$ constrained to move in a submanifold $M\subset\Tan Q$ 
defined by a family of constraints $\psi^\beta:\Tan Q\to\Real$, $\beta=1,\ldots, k$,
independent with respect to the velocities, that is
$$
\mathrm{rank}(\frac{\partial(\psi^1,\ldots,\psi^k)}{\partial(v^1,\ldots,v^n)})=k\, ,
$$
We understand that the manifold $M$ is included in $\Tan(Q\times\Real)$ and in this paragraph we assume that $M$ is a vakonomic constraint submanifold.

As a dissipative system we consider it is also subjected to the dissipative constraint given by $\psi=v_s-L=0$, and we suppose that this constraint  acts as another vakonomic constraint. Hence the constraint submanifold $B\subset M\subset\Tan(Q\times\Real)$ is defined by the family of constraints $\{\psi,\psi^\beta\}$, family which is independent with respect to the velocities $(v_s,v^1,\ldots,v^n)$ in the sense of the above rank condition.

{\bf Comment}: We can say here the same comments on the functions $\psi^\beta$ that at the beginning of Section 4, with the only difference that in this Section we understand $M\subset\Tan(Q\times\Real)$ as a vakonomic constraint submanifold and the system under study, defined by the Lagrangian $L$ and the constraints $\{\psi,\psi^\beta\}$, as a vakonomic system. In Section \ref{generalcase} we will comment the general situation with the constraints $\psi^\beta$ depending also on $s$.

\subsection{Formulation with the dynamical vector field}

As we work with a constrained vakonomic Lagrangian system, the dynamical vector field, $Y_L\in\vf(\Tan(Q\times\Real))$, is given as solution to the equation
\begin{equation}\label{vakgeneq}
\inn_{Y_L}\Omega_L-\d E_L=\mathcal{E}_{\mu\psi}+
\mathcal{E}_{\mu_\beta\psi^\beta}\, ,
\end{equation}
where $\mu,\mu_\beta:\Real\to\Real$ are the vakonomic Lagrange multipliers.
The canonical Lagrangian forms and the energy associated to $L$ are the same as in equations (\ref{lagrangforms}) and (\ref{lagrangenergy}). 

Once again as in Section 3, $\Omega_L$ is presymplectic and
$\ker\Omega_L=\displaystyle\{\derpar{}{ v_s}\}$;
hence we need to compute the primary constraints as the compatibility equation
$$
\inn_{\derpar{}{v_s}} (\inn_{Y_L}\Omega_L-\d E_L)=
\inn_{\derpar{}{v_s}} (\mathcal{E}_{\mu\psi}+
\mathcal{E}_{\mu_\beta\psi^\beta})\, .
$$
Both sides are identically zero, hence we have not any primary constraint and the equation defining the dynamical vector field is compatible all over $\Tan(Q\times\Real)$ as in the nonholonomic case.
As in the above Section 3.2, equation (\ref{vakgeneq}) develops as
$$
\inn_{Y_L}\Omega_L-\d E_L=\mathcal{E}_{\mu\psi}+
\mathcal{E}_{\mu_\beta\psi^\beta}=
\dot{\mu}\d^\mathrm{v}\psi+\mu\mathcal{E}_\psi+
\dot{\mu}_\beta\d^\mathrm{v}\psi^\beta+\sum_{\beta=1}^k\mu_\beta\mathcal{E}_{\psi^\beta} \ .
$$
In natural coordinates, the elements in the equation have the following expressions:
\begin{enumerate}
\item
For $\psi=v_s-L$,
\begin{eqnarray*}
\mathcal{E}_\psi&= &(\frac{\partial L}{\partial s}-\frac{\d}{\d t}(1))\d s-(\frac{\partial L}{\partial q^i}-\frac{\d}{\d t}(\frac{\partial L}{\partial v^i}))\d q^i \\
&=&
-\frac{\partial L}{\partial s}\d s-(\frac{\partial L}{\partial q^i}-\frac{\d}{\d t}(\frac{\partial L}{\partial v^i}))\d q^i \ , \\
\d^\mathrm{v}\psi&=&\d s-\frac{\partial L}{\partial v^i}\d q^i\, .
\end{eqnarray*} 
\item For $\psi^\beta(q,v)$,
\begin{eqnarray*}
\mathcal{E}_{\psi^\beta}&= &(\frac{\partial \psi^\beta}{\partial q^i}-\frac{\d}{\d t}(\frac{\partial \psi^\beta}{\partial v^i}))\d q^i \ , \\
\d^\mathrm{v}\psi^\beta&=&\frac{\partial \psi^\beta}{\partial v^i}\d q^i\, .
\end{eqnarray*} 
\end{enumerate}

Now in order to calculate the coordinate expression of the dynamical vector field $Y_L$, we write
$$
Y_L=A^i\frac{\partial }{\partial q^i}+B^i\frac{\partial }{\partial v^i}+
F\frac{\partial }{\partial s}+G\frac{\partial }{\partial v_s}\, ,
$$
and, in this case, we obtain the following equations:
\begin{enumerate}
\item For $\d q^i$,
\begin{eqnarray}\label{b's-vak-2}
A^h\frac{\partial^2 L}{\partial q^i\partial v^h} &-&A^h\frac{\partial^2 L}{\partial q^h\partial v^i} -B^h\frac{\partial^2 L}{\partial v^h\partial v^i}-F\frac{\partial^2 L}{\partial s\partial v^i} -v^h\frac{\partial^2 L}{\partial q^i\partial v^h} +\frac{\partial L}{\partial q^i}\nonumber\\
&=&
-\dot{\mu}\frac{\partial L}{\partial v^i}
-\mu\frac{\partial L}{\partial q^i}+
\mu(\frac{\partial^2 L}{\partial q^j\partial v^i}v^j+  \frac{\partial^2 L}{\partial v^j\partial v^i}B^j+ \frac{\partial^2 L}{\partial s\partial v^i}v_s)\nonumber\\
&+&\dot{\mu}_\beta\frac{\partial \psi^\beta}{\partial v^i}
-
\mu_\beta(\frac{\partial^2 \psi^\beta}{\partial q^j\partial v^i}v^j+  \frac{\partial^2 \psi^\beta}{\partial v^j\partial v^i}B^j)+\mu_\beta\frac{\partial \psi_\beta}{\partial q^i} \ .
\end{eqnarray}
\item For $\d v^i$,
\begin{equation}\label{reg-L-vak}
A^h\frac{\partial^2 L}{\partial v^i\partial v^h} -v^h\frac{\partial^2 L}{\partial v^i\partial v^h}
=0\, ,
\end{equation}
which implies that $A^h=v^h$ because $L$ is supposed to be regular.
\item For $\d s$,
$$
A^h\frac{\partial^2 L}{\partial s\partial v^h} -v^h\frac{\partial^2 L}{\partial s\partial v^h}+\frac{\partial L}{\partial s}=\dot\mu +\mu\frac{\partial L}{\partial s}\ ,
$$
and, being $v^h=A^h$ by equation (\ref{reg-L-vak}), this implies  
\begin{equation}\label{dotmu-2}
\dot\mu -\mu\frac{\partial L}{\partial s}=\frac{\partial L}{\partial s}\ .
\end{equation}
\end{enumerate}
The component in $\d v_s$ does not appear.
Once again, assuming that $Y_L$ must be a second order vector field in the coordinates $(q^i,v^i)$ of $\Tan Q$ and also in the dissipation coordinates $(s,v_s)$, we get $\dot{s}=v_s$
and, using the constraint $v_s-L=0$,
we obtain $F=L$. Then, introducing the values of $A^h$ and $\lambda$ in equation (\ref{b's-vak-2}),
we have that
$$
Y_L=v^i\frac{\partial }{\partial q^i}+B^i\frac{\partial }{\partial v^i}+
L\frac{\partial }{\partial s}+G\frac{\partial }{\partial v_s}\, ,
$$
where $B^i$ is obtained form equation (\ref{b's-vak}), because the Lagrangian is regular, and $D$ is still indeterminate. But, by application of the tangency condition, $\Lie_{Y_L}\phi=\Lie_{Y_L}(v_s-L)=0$, we obtain
$$
G=v^i\frac{\partial L}{\partial q^i}+B^i\frac{\partial L}{\partial v^i}+L\frac{\partial L}{\partial s}\, ;
$$
that is
$$
Y_L=v^i\frac{\partial }{\partial q^i}+B^i\frac{\partial }{\partial v^i}+
L\frac{\partial }{\partial s}+(v^i\frac{\partial L}{\partial q^i}+B^i\frac{\partial L}{\partial v^i}+L\frac{\partial L}{\partial s})\frac{\partial }{\partial v_s}\, .
$$
Observe that we need to consider the constraints $\psi^\beta$ as additional equations in order to determine the Lagrange multipliers $\mu_\alpha$ and, consequently, to obtain the coefficients $B^i$ form equation (\ref{b's-vak}). Then the dynamical vector field is completely determined except for the Lagrange multipliers. The multiplier $\mu$ is obtained form equation (\ref{dotmu-2}).
 The integral curves of the vector field $Y_L$ are the dynamical trajectories of the system.
 
 Now we obtain another expression for these trajectories in terms of the Euler-Lagrange operator.
Consider equation (\ref{b's-vak-2}) and substitute the first constraint, $\psi=v_s-L=0$,  and equations (\ref{reg-L-vak}) and (\ref{dotmu-2}), then we obtain
\begin{eqnarray*}%\label{b's-vak-2}
%A^h\frac{\partial^2 L}{\partial q^i\partial v^h} 
&-&v^h\frac{\partial^2 L}{\partial q^h\partial v^i} -B^h\frac{\partial^2 L}{\partial v^h\partial v^i}-L\frac{\partial^2 L}{\partial s\partial v^i} 
%-v^h\frac{\partial^2 L}{\partial q^i\partial v^h}
+\frac{\partial L}{\partial q^i}=\\
& &
-(1+\mu)\frac{\partial L}{\partial s}\frac{\partial L}{\partial v^i}
-\mu\frac{\partial L}{\partial q^i}+
\mu(\frac{\partial^2 L}{\partial q^j\partial v^i}v^j+  \frac{\partial^2 L}{\partial v^j\partial v^i}B^j+ \frac{\partial^2 L}{\partial s\partial v^i}v_s)=\\
& &\dot{\mu}_\beta\frac{\partial \psi^\beta}{\partial v^i}
-
\mu_\beta(\frac{\partial^2 \psi^\beta}{\partial q^j\partial v^i}v^j+  \frac{\partial^2 \psi^\beta}{\partial v^j\partial v^i}B^j)+\mu_\beta\frac{\partial \psi_\beta}{\partial q^i} \ ; %\nonumber
\end{eqnarray*}
that is
\begin{eqnarray*}
& &(1+\mu)[\frac{\partial L}{\partial q^i}-v^h\frac{\partial^2 L}{\partial q^h\partial v^i} -B^h\frac{\partial^2 L}{\partial v^h\partial v^i}- \frac{\partial^2 L}{\partial s\partial v^i}L]=\\
& &
-(1+\mu)\frac{\partial L}{\partial s}\frac{\partial L}{\partial v^i}+
\dot{\mu}_\beta\frac{\partial \psi^\beta}{\partial v^i}
+\mu_\beta(\frac{\partial \psi_\beta}{\partial q^i}-\frac{\partial^2 \psi^\beta}{\partial q^j\partial v^i}v^j-  \frac{\partial^2 \psi^\beta}{\partial v^j\partial v^i}B^j)\nonumber
\, .
\end{eqnarray*}
But, from equation (\ref{dotmu-2}) we have that
\begin{equation}\label{mu+1}
1+\mu(t)=A\exp(\int_0^t \frac{\partial L}{\partial s}\d\tau) \ ,
\end{equation}
where the integral is supposed to be defined on every curve $\gamma(t)$ solution to the dynamical equation and $A$ is a constant. Then, recalling that the coefficients $B^i$ are the components of the acceleration $\dot{v}^i$ of the curve solution, this last equation can be written as
\begin{equation*}
\mathcal{E}_L^Q=\frac{\partial L}{\partial s}\frac{\partial L}{\partial v^i}\d q^i
+\frac{1}{\mu+1}\mathcal{E}_{\mu_\beta\psi^\beta}\,.
\end{equation*}
For a curve $\gamma:\Real\to Q\times\Real$, the equation reads:
\begin{equation*}
\mathcal{E}_L^Q\circ\ddot{\gamma}=\frac{\partial L}{\partial s}\frac{\partial L}{\partial v^i}\d q^i
\circ\dot{\gamma}
+A^{-1}\exp(-\int_0^t \frac{\partial L}{\partial s}\d\tau)\mathcal{E}_{\mu_\beta\psi^\beta}\circ\ddot{\gamma}\,.
\end{equation*}
Observe that the constant $A$ can not be zero as we have proved in a similar situation with equation (\ref{A-mu+1}).

We must add the constraints $\psi^\beta=0$ to this equation  in order to solve for the dynamical trajectories $\gamma(t)=(q^i(t),s(t),\mu_\beta(t))$. Observe that, as this equation contains the derivatives of the Lagrange multipliers, we need to include their initial values as initial conditions to integrate the differential equations. The same comment is valid for the multiplier $\mu$ and the constant $A$.

This is the classical equation for mechanical systems with vakonomic constraints but with the term that corresponds to the dissipative situation, the first term of the left hand side,  and with the terms giving account to the dissipation on the constraints along the evolution of the system.

If the Lagrangian $L$ does not depend on the dissipation variable $s$, then we obtain the classical equation for vakonomic systems.

\subsection{Formulation with the Euler-Lagrange operator}

In this approach the dynamical equation is given by
$$
\mathcal{E}_L=\mathcal{E}_{\mu\psi}+
\mathcal{E}_{\mu_\beta\psi^\beta}=
\dot{\mu}\d^\mathrm{v}\psi+\mu\mathcal{E}_\psi+
\dot{\mu}_\beta\d^\mathrm{v}\psi^\beta+\mu_\beta\mathcal{E}_{\psi^\beta}\,.
$$
If we apply the usual decomposition for $Q\times\Real$ we have:
\begin{equation*}
\mathcal{E}_L^Q+\frac{\partial L}{\partial s}\d s=-\dot\mu\frac{\partial L}{\partial v^i}\d q^i+
\dot\mu\d s-\mu\mathcal{E}_L^Q
-\mu\frac{\partial L}{\partial s}\d s+\dot\mu_\beta\d^\mathrm{v}\psi^\beta+\mu_\beta\mathcal{E}_{\psi^\beta}\,.
\end{equation*}
Once again identifying the terms with $\d s$ and $\d q^i$ in both sides of the equation, we obtain two equations. The first one is
\begin{equation}\label{dotmu-mu+1}
\dot\mu=(1+\mu)\frac{\partial L}{\partial s}\,,
\end{equation}
and substituting this last expression into the other,
\begin{equation}\label{mu+1-EL}
(\mu+1)\mathcal{E}_L^Q=-(\mu+1)\frac{\partial L}{\partial s}\frac{\partial L}{\partial v^i}\d q^i
+\dot\mu_\beta\d^\mathrm{v}\psi^\beta+\mu_\beta\mathcal{E}_{\psi^\beta}\,.
\end{equation}
But, as above, in equation (\ref{mu+1}) we have that
\begin{equation}\label{vak-eq}
1+\mu(t)=A\exp(\int_0^t \frac{\partial L}{\partial s}\d\tau)\,.
\end{equation}
Hence, for a dynamical trajectory of the system $\gamma:\Real\to Q\times\Real$ we have
obtained
\begin{equation*}
\mathcal{E}_L^Q\circ\ddot{\gamma}=\frac{\partial L}{\partial s}\frac{\partial L}{\partial v^i}\d q^i
\circ\dot{\gamma}
+A^{-1}\exp(-\int_0^t \frac{\partial L}{\partial s}\d\tau)\mathcal{E}_{\mu_\beta\psi^\beta}\circ\ddot{\gamma}\,,
\end{equation*}
which can be written as
\begin{equation*}
\mathcal{E}_L^Q\circ\ddot{\gamma}=\frac{\partial L}{\partial s}\d^\mathrm{v}L
\circ\dot{\gamma}
+A^{-1}\exp(-\int_0^t \frac{\partial L}{\partial s}\d\tau)\mathcal{E}_{\mu_\beta\psi^\beta}\circ\ddot{\gamma}\ ,
\end{equation*}
and we need to add the constraints to have a system with enough equations to determine the components of the dynamical trajectories and the Lagrange multipliers. As usual, in vakonomic problems, we need to give initial conditions for the multipliers as they and their derivatives are included in the final dynamical equations.

\subsection{The general case for constraints}
\label{generalcase}

Suppose now that the constraints $\psi^\alpha$ depend on $s$, the dissipation variable, then $\psi^\alpha=\psi^\alpha(q^i,v^i,s)$ and we have 
$$
\d^\mathrm{v}\psi^\beta=\d_Q^\mathrm{v}\psi^\beta
\quad ,\quad
\mathcal{E}_{\psi^\beta}=\mathcal{E}_{\psi^\beta}^Q+\frac{\partial \psi^\beta}{\partial s}\d s \ ,
$$
and separating the components in $\d q^i$ and $\d s$, as above, we obtain, instead of the differential equation (\ref{dotmu-mu+1}), the following one for the multiplyer $\mu$
\begin{equation}
\label{dotmu-general2}
\dot\mu=(1+\mu)\frac{\partial L}{\partial s}+\mu_\beta\frac{\partial \psi^\beta}{\partial s}
\,.
\end{equation}
Thus equation(\ref{mu+1-EL}) transforms into
\begin{equation}\label{full_vakonomic}
(\mu+1)\mathcal{E}_L^Q=-(\mu+1)\frac{\partial L}{\partial s}\frac{\partial L}{\partial v^i}\d q^i
+\mu_\beta\frac{\partial \psi^\beta}{\partial s}\frac{\partial L}{\partial v^i}\d q^i
+\dot\mu_\beta\d_Q^\mathrm{v}\psi^\beta+\mu_\beta\mathcal{E}^Q_{\psi^\beta}\ ;
\end{equation}
that is,
$$
(\mu+1)\mathcal{E}_L^Q=-(\mu+1)\frac{\partial L}{\partial s}\d_Q^\mathrm{v}L
+\mu_\beta\frac{\partial \psi^\beta}{\partial s}\d_Q^\mathrm{v}\psi^\beta
+\mathcal{E}^Q_{\mu_\beta\psi^\beta}\,,
$$
which is a new equation, different from (\ref{vak-eq}). The second term on the right hand side gives account of the dependence of the constraints on $s$. With respect to the multiplyer $\mu$, we can obtain an integral expression from the linear differential equation (\ref{dotmu-general2}).

Hence in this case the corresponding dynamical equations change.

\subsection{The reduced equation}

If $(\mu+1)$ does not vanish, which, according to~\eqref{vak-eq}, is always the case when the constraints do not depend on $s$, the equation can be reduced. We define the new multipliers
$$
    \nu^\beta = \frac{\mu^\beta}{\mu + 1} \ .
$$
Dividing~\eqref{full_vakonomic} by $(\mu+1)$, we obtain
\begin{equation}
\label{reduced_vakonomic_0}
\mathcal{E}_L^Q=-\frac{\partial L}{\partial s}\frac{\partial L}{\partial v^i}\d q^i
+\nu_\beta\frac{\partial \psi^\beta}{\partial s}\frac{\partial L}{\partial v^i}\d q^i
+ \frac{\dot{\mu}_\beta}{\mu + 1} \d_Q^\mathrm{v}\psi^\beta+\nu_\beta\mathcal{E}^Q_{\psi^\beta} \ .
\end{equation}
Using the expression~\eqref{dotmu-general2}, we compute
$$
    \dot{\nu}_\beta = \frac{\dot{\mu}_\beta}{\mu + 1} - (\frac{\partial L}{\partial s} + \mu_\alpha \frac{\partial \psi^\alpha}{\partial s} ) \nu_\beta \ .
$$
Substituting on~\eqref{reduced_vakonomic_0}, we obtain
$$
%\label{reduced_vakonomic_1}
\mathcal{E}_L^Q=-\frac{\partial L}{\partial s}\frac{\partial L}{\partial v^i}\d q^i
+\nu_\beta\frac{\partial \psi^\beta}{\partial s}\frac{\partial L}{\partial v^i}\d q^i
+ (\dot{\nu}_\beta + (\frac{\partial L}{\partial s} + \mu_\alpha \frac{\partial \psi^\alpha}{\partial s} ) \nu_\beta ) \d_Q^\mathrm{v}\psi^\beta+\nu_\beta\mathcal{E}^Q_{\psi^\beta} \ ;
$$
that is,
$$
    \mathcal{E}_{L - \nu_\beta \psi}^Q=(-\frac{\partial L}{\partial s} + \nu_\alpha\frac{\partial \psi^\alpha}{\partial s} )\d_Q^\mathrm{v}( L - \nu_\beta \psi^\beta ) \ ,
$$
and this equation is the one derived on~\cite{LLM-2021}.

\subsection{Summary of Section 5}

We have studied the dynamics of a dissipative Lagrangian $\mathfrak{L}\colon\Tan Q\times\Real\to\Real$, $\mathfrak{L}(q^i,v^i,s)$, as a dissipative system and constrained to a submanifold $M\subset\Tan Q$
defined by a family of constraints $\psi^\beta:\Tan Q\to\Real$, $\alpha=1,\ldots, k$,  
independent with respect to the velocities $(v^1,\ldots,v^n)$, as a vakonomic system given by the Lagrangian system $(\Tan(Q\times\Real),L=\pi_o^*\mathfrak{L})$ constrained by the family 
$(\psi(q^i,v^i,s,v_s)=v_s-L=0, \psi^\beta)$ and we have obtained the following

\begin{teor}
The dissipative dynamics corresponding to the dissipative Lagrangian system $(\Tan(Q\times\Real),L=\pi_o^*\mathfrak{L})$ constrained by the family $\psi^\alpha:\Tan Q\to\Real$, $\alpha=1,\ldots, k$, of vakonomic constraints, is given, 
for a dynamical trajectory of the system $\gamma:\Real\to Q\times\Real$, by the equation 
\begin{equation*}
\mathcal{E}_L^Q\circ\ddot{\gamma}=\frac{\partial L}{\partial s}\frac{\partial L}{\partial v^i}\d q^i
\circ\dot{\gamma}
+A^{-1}\exp(-\int_0^t \frac{\partial L}{\partial s}\d\tau)\mathcal{E}_{\mu_\beta\psi^\beta}\circ\ddot{\gamma}\,,
\end{equation*}
which can be written, in a more intrinsic expression, as
\begin{equation*}
\mathcal{E}_L^Q\circ\ddot{\gamma}=\frac{\partial L}{\partial s}\d^\mathrm{v}L
\circ\dot{\gamma}
+A^{-1}\exp(-\int_0^t \frac{\partial L}{\partial s}\d\tau)\mathcal{E}_{\mu_\beta\psi^\beta}\circ\ddot{\gamma}\ ,
\end{equation*}
and we need to add the constraints to have a system with enough equations to determine the components of the dynamical trajectories and the Lagrange multipliers. As usual, in vakonomic problems, we need to give initial conditions for the multipliers as they and their derivatives are included in the final dynamical equations. The same situation is for the constant $A$.

\end{teor}

\section{Conclusions and outlook}

%In this paper we have developed an unified %description of  nonholonomic and vakonomic %dynamics for the so-called dissipative or %contact Lagrangian systems subject to %constraints. 

In this paper we have obtained the Herglotz equations for a dissipative Lagrangian as a classical Lagrangian system, with a dissipative variable $s$, but as a constrained Lagrangian system, both as nonholonomic and vakonomic approach, instead of using the Herglotz variational principle. In that way we obtain the classical contact equations for dissipative systems.

Furthermore we have developed the general theory of dissipative Lagrangian systems subject to constraints with
a full description of  nonholonomic and vakonomic dynamics. The equations we obtain reduce to the classical constrained ones in the case that the Lagrangian does not depend on the dissipation variable.

The tools are just the presymplectic geometry and the Euler-Lagrange operator. 
This approach allows us to have a clear distinction when the constraints for our system are just $\dot{s} = L$ or we have additional constraints.

From these results we can consider a series of problems to be discussed further, among them:

- The potential applications to the theory of optimal control of dissipative systems, driven by an appropriate Pontriaguin Maximal Principle for the contact case. This study would permit us to compare the results with those obtained in a previous work \cite{LLM-2021b}.

- The study of symmetries and the corresponding dissipated quantities. 
This will allow us a comparison with the results obtained in
this paper an in other previous papers for the case of nonholonomic and vakonomic systems in the symplectic context
\cite{DeLeon2020,GGMRR-2019b,MCL-2000}.

- To discuss a comparison among the solutions of the nonholonomic and vakonomic equations in the same vein that for the symplectic case
(see \cite{CLMM-2002,LMM,GMM-2003} and the results in this paper).

%%%%%%%%%%%%%%%%%%%%%%%%%%%%%%%%%%%%%%%%%%%%%%%%%%%%%%%%%%%%%%%%

\section*{Acknowledgments}

We acknowledge the financial support from the Spanish
Ministerio de Ciencia, Innovaci\'on y Universidades project
PGC2018-098265-B-C33;
the MINECO Grant MTM2016-76-072-P; 
PID2019-106715GB-C21 Grant and “Severo Ochoa Programme for Centres of Excellence in R$\&$D” (CEX2019-000904-S).
Manuel La\'inz wishes to thank MICINN and ICMAT for a FPI-Severo Ochoa predoctoral contract PRE2018-083203.
We also thank the referee for his careful reading of the manuscript and his comments that have allowed us to improve some parts of the work.

%%%%%%%%%%%%%%%%%%%%%%%%%%%%%%%%%%%%%%%%%%%%%%%%%%%%%%%%%%%%%%%%

\end{document}